\documentclass{article}
\usepackage{etoolbox}
\newtoggle{arxiv}
\toggletrue{arxiv}
\usepackage{packages-preamble}
\usepackage{wrapfig}
\usepackage{diagbox}
\usepackage{subcaption}
\newif\ifcomments

\commentsfalse
\usepackage{karan-macros}
\title{Auditing Private Prediction}

\author{Karan Chadha \footnote{Part of this work was done during a Google DeepMind internship.} \\ Stanford University \and Matthew Jagielski \\ Google DeepMind \and Nicolas Papernot \\ Google DeepMind \and Christopher Choquette-Choo \\ Google DeepMind \and Milad Nasr \\ Google DeepMind}
\date{\today}

\begin{document}

\maketitle

\begin{abstract}
Differential privacy (DP) offers a theoretical upper bound on the potential privacy leakage of an algorithm, while empirical auditing establishes a practical lower bound. 
Auditing techniques exist for DP training algorithms. However machine learning can also be made private at inference. %
We propose the first framework for auditing private prediction where we instantiate adversaries with varying poisoning and query capabilities. This enables us to study the privacy leakage of four private prediction algorithms:  PATE~\citep{papernot2016semi}, CaPC~\citep{choquette2020capc}, PromptPATE~\citep{duan2023flocks}, and Private-kNN~\citep{zhu2020private}.
To conduct our audit, we introduce novel techniques to empirically evaluate privacy leakage in terms of Renyi DP.
Our experiments show that (i) the privacy analysis of private prediction can be improved, (ii) algorithms which are easier to poison lead to much higher privacy leakage, and (iii) the privacy leakage  is significantly lower for adversaries without query control than those with full control.

\end{abstract}

\section{Introduction}

Differential privacy (DP) assesses an algorithm's privacy by examining its outputs on two adjacent datasets, $\trainset$ and $\trainset'$, which differ in one data point~\citep{dwork2006calibrating}. 
It bounds the log ratio of output distribution probabilities on these datasets using a parameter $\diffp$. 
A small $\diffp$ ensures that an adversary cannot confidently distinguish whether the algorithm processed $\trainset$ or $\trainset'$.
Thus, $\diffp$ analytically bounds private information leakage from the algorithm's outputs. 
In contrast, \textit{auditing} a private algorithm~\cite{ding2018detecting, jagielski2020auditing} provides a lower bound on its privacy leakage.
Analyzing both upper and lower bounds can yield three insights:
a large gap may indicate a potential slack in the analysis; a lower bound surpassing the upper may indicate an incorrect analysis or implementation~\cite{tramer2022debugging}; and tracking how these bounds shift with changes to assumptions on the adversary's capabilities and knowledge can inform us of which assumptions contribute most to the algorithm's privacy leakage~\cite{nasr2021adversary}.

In the context of machine learning (ML), existing work has exclusively audited differentially private \textit{training} algorithms. These algorithms train models satisfying DP~\citep{abadi2016deep}, ensuring the privacy of all model predictions due to DP's post-processing property.\footnote{An arbitrary function can be applied to the outputs of a DP algorithm with no consequences to privacy.} However, machine learning can also be made private at \textit{inference}. Here, models are trained non-privately and their predictions are noised before release to satisfy DP.
Despite the increasing relevance of private prediction, notably to task adaptation of large language models~\citep{duan2023flocks}, there are no known techniques to audit such algorithms. Our work addresses this gap and proposes the first auditing framework to do so.

Private prediction algorithms diverge from private training by training multiple non-private teacher models on separate data partitions. At inference, they compile individual model predictions into a histogram, introduce noise to each histogram bin, and select the most frequent bin as the output.
To quantify the privacy leakage of private prediction, we  upper bound several cumulants of the log ratio of output distributions (instead of the log ratio directly), resulting in a Renyi DP (RDP) guarantee. RDP, an alternative formulation of DP, offers enhanced compositional properties which is beneficial when assessing privacy leakage that is composed across multiple test queries.
When reporting, we convert the composed RDP guarantees to a classical DP guarantee. 

To audit private prediction, the standard approach would consider the full output set of multiple queries and audit the resulting distribution using classical auditing techniques. However, the discrete high dimensional nature of the output distribution makes the application of these techniques non-trivial. Moreover, composition theorems in classical DP analysis are lossy and composing corresponding lower bounds is invalid. This necessitates developing a novel methodology to audit RDP guarantees which we use to audit per-query privacy leakage and further compose using lossless RDP composition theorems to give lower bounds on the privacy leakage across multiple queries. Along with our new approach to RDP auditing, we also introduce a technique to calculate the exact Renyi divergence between the outputs of noisy argmax on neighboring histograms. This dual approach enables a more nuanced assessment and helps in attributing discrepancies between the audit and the theoretical analysis to either looseness in the analysis or strength of the adversary, thereby enhancing the clarity of our audit.

We apply our auditing framework to four well-known private prediction algorithms: PATE \citep{papernot2016semi,papernot2018scalable}, CaPC \citep{choquette2020capc}, PromptPATE \citep{duan2023flocks} and Private-kNN \citep{zhu2020private}. All of these algorithms except Private-kNN allow for a data-dependent privacy analysis, where the standard data-independent analysis can be refined in cases when most models agree on the prediction to be made. Even so, our exact privacy analysis is tighter than this data-dependent calculation indicating potential slack in the analysis. Across all algorithms, we find that an adversary's capability to control the test queries is an important factor contributing to maximal privacy leakage. Furthermore, more privacy is leaked in cases where it is easier for adversaries to impact the behavior of a teacher model on multiple test queries. These trends are highlighted not only by the leakage due to different adversaries, but also due to the differences in the relative leakage of different private prediction algorithms correlating with their ease of poisoning.

\subsection{Related Work}
\citet{malek2021antipodes} run attacks on PATE under label DP (a variant of DP protecting just the labels), for a specific type of adversary. Our goal is to audit private prediction in general, and with more granularity to enable improved attacks or analysis. \citet{wang2022differential} show \emph{test-time} attacks on the queries to private prediction, while our goal is to measure privacy leakage of the training data. 

\citet{ding2018detecting} propose an approach to detect DP violations in many classic DP algorithms including noisy argmax, which we consider in our work. Our work requires RDP auditing rather than their $\epsilon$-DP auditing, and we audit using inputs that are relevant to private prediction, to measure the privacy leakage from these ML algorithms rather than just the noisy argmax.

Our work measures leakage of private prediction algorithms through their returned labels. \citet{choquette2021label, li2021membership} demonstrate membership inference attacks on traditional classifiers using only returned labels.

Recent work on auditing has shown how to lower bound general RDP guarantees using variational characterizations \cite{kong2023r}; our work instead performs audits suited for mechanisms with a discrete output space and with an attack interpretation. See Appendix~\ref{sec:related-works} for full details.

\section{Notation and Preliminaries}\label{sec:prelim}

We work in the multiclass classification setting where we use features from $\mc{X} \subset \R^d$ to predict labels from $\mc{Y} = {1,\dots,\numclasses}$. Let the space of feature vector and label pairs be $\trainspace = \mc{X}\times\mc{Y}$. 
{Let $x \in \mc{X}$ and $y \in \mc{Y}$, with $\trainset = (x_i,y_i)_{i = 1}^{n}$ the training dataset. }
In ML, our task is to learn a function (model) $f(\cdot;\trainset): \mc{X} \to \mc{Y}$ using the training dataset $\trainset$, such that $f(x;\trainset)$ accurately estimates the label $y$ of the feature $x$. 
{A long line of work has shown that machine learning models can leak sensitive information about its training dataset~\citep{shokri2016membership,tramer2022truth,carlini2021extracting}.
The gold standard for preventing this is differential privacy (see Appendix~\ref{sec:related-works} for more details).
}

\subsection{Differential Privacy}
\label{ssec:dp}

\begin{definition}[Approximate DP]\citep{dwork2006calibrating}\label{definition:differential-privacy}
    An algorithm $\mech : \trainspace^n \to \mc{O}$ is {${\adiffp}$-DP} 
    if for any measurable $O \in \mc{O}$ and $\trainset,\trainset' \in \trainspace^n$ that differ in one entry, we have
    $\P\left(\mech(\trainset) \in O\right) \leq e^{\diffp} \cdot \P\left(\mech(\trainset') \in O\right) + \delta.
    $
\end{definition}
{An important property of DP is sequential composition: given two mechanisms $\mech_1,\mech_2$ each satisfying a finite DP bound, $\mech_2(\trainset,\mech_1(\trainset))$ also has a finite DP bound. However, the composition bounds for approximate DP are not exact and lead to loose privacy accounting.}
{Renyi DP (RDP) is an alternative definition to 
$\adiffp$-DP that provides lossless composition bounds. Thus, RDP is the analysis method of choice for private prediction algorithms. The Renyi divergence between distributions $P$ and $Q$ is defined as:}
\begin{equation*}
    D_\alpha(P || Q) \coloneqq \frac{1}{\alpha - 1}\log \E_{x \sim P}\brc{\frac{P(x)}{Q(x)}^\alpha}.
\end{equation*}

\begin{definition}[Renyi DP]\citep{mironov2017renyi}
    An algorithm $\mech : \trainspace^n \to \mc{O}$ satisfies $(\alpha,\rdp{\alpha})$-RDP, if for any $\trainset, \trainset' \in \trainspace^n$ that differ in one entry we have
    $D_\alpha(\mech(\trainset) || \mech(\trainset')) \leq \rdp{\alpha}.
    $
\end{definition}

The RDP guarantee for the composition of mechanisms is simply the order-wise sum of individual RDP guarantee. We keep track of the RDP for several orders, and report the optimal $\adiffp$-DP guarantee across orders found using the following theorem.
\begin{theorem}\citep[Thm. 20]{balle2020hypothesis}
\label{th:balle}
    If an algorithm is $(\alpha,\rdp{\alpha})$-RDP, then it is also $(\rdp{\alpha} + \log(\frac{\alpha - 1}{\alpha}) - \frac{\log{\delta} + \log{\alpha}}{\alpha - 1},\delta)$-DP for any $\delta \in (0,1)$.
\end{theorem}

\subsection{Auditing Differential Privacy}
Unlike the upper bounds that any DP algorithm must satisfy, auditing gives a lower bound.
We define \textit{auditing} the approximate DP guarantees of an algorithm $\mech$ as empirically estimating parameters $\lb{\diffp}$ and $\delta$ such that 
the algorithm does not satisfy $\adiffp$-DP for any $\diffp<\lb{\diffp}$. We denote it as $\lb{\diffp}$ since it is a lower bound on the true $\adiffp$-DP satisfied by the algorithm. To find such a lower bound for an algorithm $\mech$, we choose neighbouring datasets $\trainset$, $\trainset'$ and an output set $O$ on which we run the algorithm $\mech$ on $\trainset$ and $\trainset'$ many times (say $\numexp$) and check the number of times the output is in $O$ for both $\trainset$ and $\trainset'$ respectively. Using these proportions and the Clopper-Pearson confidence intevals \citep{clopper1934use}, we calculate upper and lower bounds on the probabilities $\P\left(\mech(\trainset) \in O\right)$ (denoted as $p_0^\ell,p_0^u$) and $\P\left(\mech(\trainset') \in O\right)$ (denoted as  $p_1^\ell,p_1^u$) and subsequently estimate the lower bound $\lb{\diffp}$ using the following equation:
\iftoggle{arxiv}{}{
\vspace{-2mm}
}

\begin{equation}\label{eq:lb-adiffp}
    \lb{\diffp} = \max\brc{\frac{p_1^\ell - \delta}{p_0^u},\frac{p_0^\ell - \delta}{p_1^u}}.
\end{equation} 
The choice of the neighbouring datasets and the output set  partly hinges on the constraints imposed on an adversary, influencing the strength of the calculated lower bound $\lb{\diffp}$. 

\iftoggle{arxiv}{\paragraph{Attack Interpretation.}}{\textbf{Attack interpretation.}}
We can also view auditing as a hypothesis test, where
the adversary resolves:
\iftoggle{arxiv}{}{
\vspace*{-2mm}
}
\begin{align*}
        &H_0: \trainset \text{ was used to generate the output of the algorithm } \mech  \\
        &H_1: \trainset' \text{ was used to generate the output of the algorithm } \mech,
\end{align*}
using the output of the mechanism. For a rejection region $O$, let the false positive ($\P(\mech(\trainset) \in O)$) and false negative ($\P(\mech(\trainset') \in O')$) rates for this hypothesis test be FP and FN respectively. Then, \Cref{eq:lb-adiffp} can be reformulated using upper bounds on FP and FN as:
\iftoggle{arxiv}{}{
\vspace*{-2mm}
}
\begin{equation*}
    \lb{\diffp} = \max\brc{\frac{1 - \delta - FN^u}{FP^u},\frac{1 - \delta - FP^u}{FN^u}}.
\end{equation*}

\section{Private Prediction}

\iftoggle{arxiv}
{
\begin{algorithm}
\caption{Private prediction framework}\label{alg:prediction-framework}
    \DontPrintSemicolon
    \SetKwInOut{Input}{Input}
    \SetKwInOut{Output}{Output}
    \SetKwInOut{Params}{Params}
    \Input{Training dataset $\trainset=(X, Y)$, teacher count $\numteachers$, query dataset $\queryset$, per-query noise scale $\sigma$}
    \Params{Number of partitions $m$}

    $\textsc{Predictions} = []$\;
    \textbf{Training Phase:}\;
    $P = \textsc{Partition}(\trainset, m)$ \Comment{Partition into $m$ splits}\;

    \For{$i\in [1, m]$}{
    $T_i = \textsc{TrainTeacher}(P_i)$\;
    }

    \textbf{Prediction Phase:}\;
    \For{$x \in \queryset$}{
    $T = \textsc{Choose}(T_1^m,k,x)$  \Comment{Choose $\numteachers$ teachers from $T_1^m$} \;
    $\counts{\trainset} = \textsc{GetPredictions}(T, x)$   \Comment{Get predictions from $\numteachers$ chosen teachers} \;
    $\noisycounts{\trainset} = \textsc{AddNoise}(\counts{\trainset}, \sigma) \label{line:noise}$\;
    $\textsc{Predictions}.append(\text{argmax } \noisycounts{\trainset}$)\;
    }
    \Return $\textsc{Predictions}$
\end{algorithm}
}{

\begin{algorithm}[tb]
   \caption{Private prediction framework}
   \label{alg:prediction-framework}
\begin{algorithmic}[1]
   \STATE {\bfseries Input:}Training dataset $\trainset=(X, Y)$, teacher count $\numteachers$, query dataset $\queryset$, per-query noise scale $\sigma$
   \STATE {\bfseries Params:} Number of partitions $m$
   \STATE Initialize $\textsc{Predictions} = []$
   \STATE \textbf{Training Phase:}
   \STATE $P = \textsc{Partition}(\trainset, m)$ 
   \FOR{$i=1$ {\bfseries to} $m$}
   \STATE $T_i = \textsc{TrainTeacher}(P_i)$
   \ENDFOR
   \STATE \textbf{Prediction Phase:} 
   \FOR{$q \in \queryset$}
  \STATE $T = \textsc{Choose}(T_1^m,k,q)$  
   \STATE $\counts{\trainset} = \textsc{GetPredictions}(T, q)$   
   \STATE $\noisycounts{\trainset} = \textsc{AddNoise}(\counts{\trainset}, \sigma) \label{line:noise}$
   \STATE $\textsc{Predictions}.append(\text{argmax } \noisycounts{\trainset}$)
   \ENDFOR
\end{algorithmic}
\end{algorithm}
}
In this section, we define privacy-preserving prediction, integrate known algorithms into a unified framework, and present two prevalent methods for reporting the privacy guarantee in this framework. Following the setup of \cite{dwork2018privacy}, let $\mech(\trainset)$ denote a prediction interface produced by applying algorithm $\mech$ to the training set $\trainset$,  which produces an output in $\mc{Y}$ when queried with a point in $\mc{X}$. Let $(\mech(\trainset) \leftrightarrow \queryset)$ denote the sequence of outputs of the interface when queried with  a test query sequence $\queryset$.
\begin{definition}[Private Prediction Interface]\label{def:priv-pred-interface}
    A prediction interface $\mech(\trainset)$ 
    satisfies $(\alpha,\rdp{\alpha})$-RDP if for every test query sequence $\queryset$, the output $(\mech(\trainset) \leftrightarrow \queryset)$
    satisfies $(\alpha,\rdp{\alpha})$-RDP with respect to dataset $\trainset$.
\end{definition}
\Cref{alg:prediction-framework} gives a general framework for private prediction, based on the Sample Aggregate framework \citep{nissim2007smooth}. It first divides the training dataset into $m$ splits, training a distinct teacher on each split. 
In the prediction phase, for each query, it selects $\numteachers$ teachers from the $m$ available, aggregating their predictions on the query point $q$ into a histogram $\counts{\trainset}$. After adding noise to create $\noisycounts{\trainset}$, the label that achieves plurality, i.e., the label with the highest count in $\noisycounts{\trainset}$, becomes the prediction for $q$. The added noise typically follows a Gaussian distribution with scale $\sigma$, chosen to satisfy the desired privacy guarantee. Mathematically,
\begin{equation} \label{eq:noisy-argmax}
    \textsc{Predictions} = \argmax_{y \in \mc{Y}} \{\counts{\trainset} + \normal(0,\sigma^2)\}.
\end{equation}
The privacy guarantee of \Cref{alg:prediction-framework} hinges on the noisy argmax (\cref{eq:noisy-argmax}) mechanism's privacy properties.
We analyze the privacy for each query individually using RDP and combine them using composition theorems. Working with RDP offers tighter composition properties than approximate DP and it facilitates data-dependent privacy leakage analysis (see below).
The RDP guarantee is calculated in two ways:

\iftoggle{arxiv}{
\paragraph{Data Independent.}
}{
\textbf{Data Independent.}
}
This approach overlooks the histogram's post-processing into a single label release and assumes the release of the entire noisy histogram. It results in conservative privacy accounting, assuming a worst-case scenario for each query and disregards any privacy amplification due to post-processing.

\iftoggle{arxiv}{
\paragraph{Data Dependent.}
}{
\textbf{Data Dependent.}
} 
Proposed in \citet[Appendix A]{papernot2018scalable}, this accounting technique applies data-dependent analysis whereby the prediction interface incurs a smaller privacy cost when many teachers agree on a label. It partly accounts for post-processing for these easy queries, which improves accounting.  In this case, the privacy parameter itself may release some private information. However, we can use the smooth sensitivity mechanism \citep[Appendix B]{papernot2018scalable} to release it privately.

\section{Auditing Private Prediction}

\begin{figure}[t]
    \centering
    \includegraphics[width=\columnwidth]{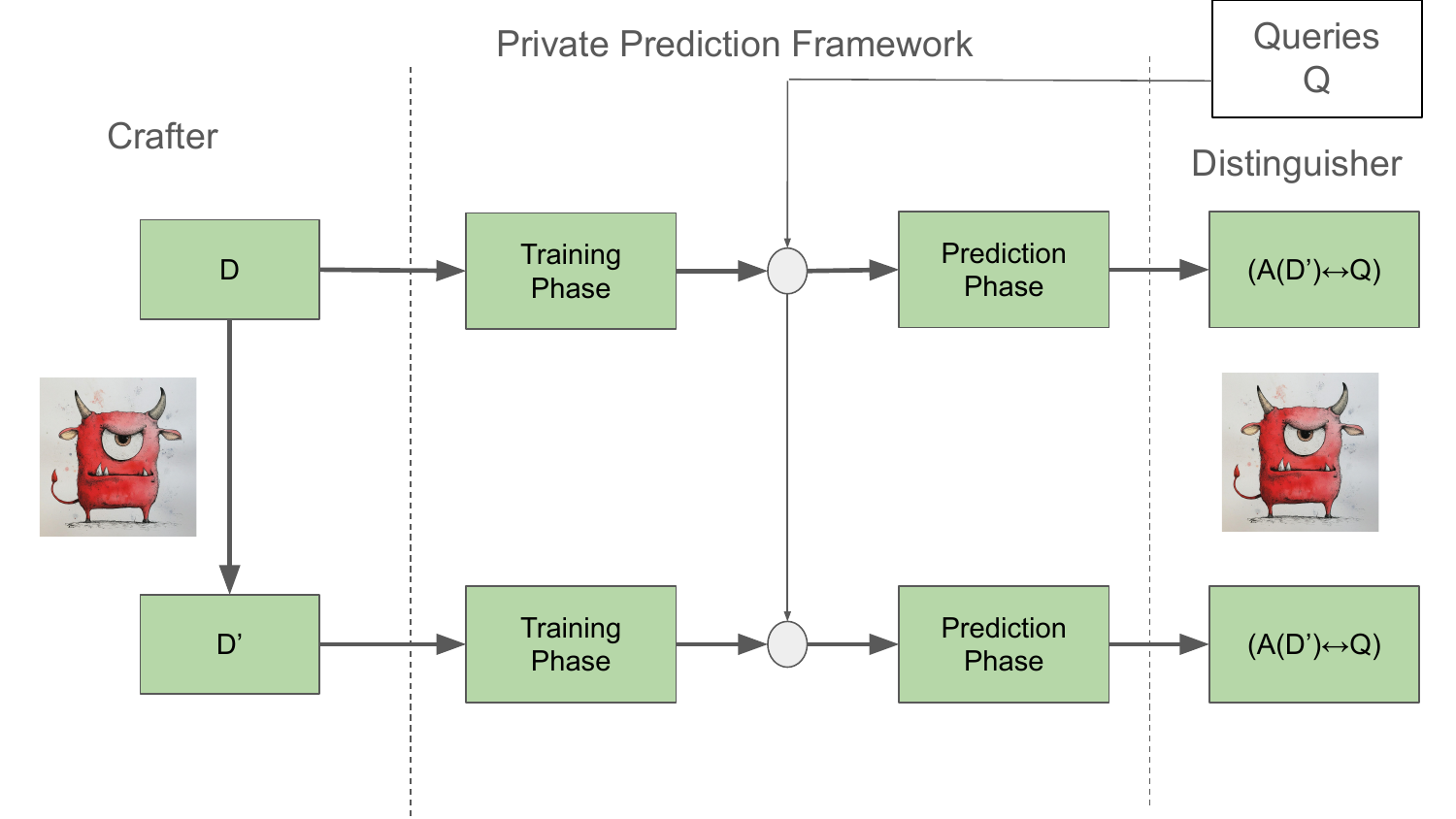}
    \caption{Framework to audit private prediction algorithms.}
    \label{fig:audit_priv_pred_framework}
\end{figure}

We now present our framework for auditing private prediction. 
Like \citet{nasr2021adversary}, we follow an \emph{adversary instantiation} approach; we define different adversaries based on their training data altering capabilities and the freedom to choose the test queries. 
Then, we discuss how we use the results of an adversary's attack as input to  audit the RDP guarantees provided by the noisy argmax, which is the core privacy primitive in private prediction. 
\Cref{fig:audit_priv_pred_framework} shows our framework, which we divide into three parts. 

\begin{enumerate}[leftmargin=4mm]
    \setlength\itemsep{0em}
    \item \textit{Crafter.} Starting with a dataset $\trainset$, the crafter generates a new dataset $\trainset'$, differing from $\trainset$ by a single data point. These datasets then become the training inputs for private prediction algorithms. We recognize two types of crafters based on their methods of constructing $\trainset'$ from $\trainset$:
    \begin{itemize}[leftmargin=4mm]
        \item Natural Crafter: Adds an in-distribution point.
        \item Poisoning Crafter: Adds an adversarial point.
    \end{itemize}
    \item \textit{Private Prediction Algorithm.} The private prediction framework $\mech(\cdot)$, detailed in \Cref{alg:prediction-framework}, accepts datasets $\trainset$ and $\trainset'$ to create private prediction interfaces $\mech(\trainset)$ and $\mech(\trainset')$, by passing the datasets through the \textit{Training Phase}. These interfaces then respond to a sequence of queries $\queryset = (q_1,q_2,\dots,)$ with sequences $(\mech(\trainset) \leftrightarrow \queryset) = (\mech(\trainset;q_1),\mech(\trainset;q_2),\dots,)$ and $(\mech(\trainset') \leftrightarrow \queryset) = (\mech(\trainset';q_1),\mech(\trainset';q_2),\dots,)$. We audit PATE, CaPC, PromptPATE and Private-kNN.
    \item \textit{Distinguisher.} The distinguisher chooses the query sequence  $\queryset$, observes the output of the interfaces $(\mech(\trainset) \leftrightarrow \queryset)$ and $(\mech(\trainset') \leftrightarrow \queryset)$, and estimates the privacy leakage of the prediction framework by comparing the output distributions. We study two distinguishers based on their query selection methods:
    \begin{itemize}[leftmargin=3mm]
        \item Natural Distinguisher: Chooses queries $(\queryset)$ from a natural distribution, mimicking real-world scenarios, simulated using the test dataset.
        \item Adversarial Distinguisher: Chooses $\queryset$ adversarially, which, in all cases we consider, is querying the interface with the same query $q$ repeatedly.
    \end{itemize}
\end{enumerate}

\subsection{Adversaries}
Using the above framework, we define three adversaries using the aforementioned crafters and the distinguishers.

\iftoggle{arxiv}{
\paragraph{$\natmiadvquer$.}
}{
$\natmiadvquer$.
}

Natural crafter, adversarial distinguisher, simulating a membership inference adversary who tries to infer the membership of an in-distribution training set example, but who adversarially chooses queries.

\iftoggle{arxiv}{
\paragraph{$\poismiadvquer$.}
}{
$\poismiadvquer$.
} Poisoning crafter, adversarial distinguisher, simulating a stronger adversary with the power to choose a worst case poisoning point to maximize the efficacy of a membership inference attack. 

\iftoggle{arxiv}{
\paragraph{$\poisminatquer$.}
}{
$\poisminatquer$.
} Poisoning crafter, natural distinguisher, where the adversary statically poisons $\trainset$ to form $\trainset'$, but can only use natural queries to distinguish.
The responses of a private prediction interface on such queries facilitate training a student model (as suggested in the original formulation of PATE, CaPC and PromptPATE), which can then be used to answer queries indefinitely without any additional privacy cost. The privacy leakage of the student model is bounded by the privacy leakage of $\poisminatquer$ due to data processing inequality, though student training may not reduce leakage~\citep{jagielski2023students}. The restriction to natural queries is important since with a reasonably small privacy budget, a performant student model can only be trained on a natural set of queries.

\subsection{Auditing RDP of Noisy Argmax}\label{sec:audit-rdp}

The standard approach to auditing would take the full output of the private prediction algorithm on the entire set of queries, and perform an attack using this sequence of outputs. However, the space of outputs is very high dimensional: with $\numqueries$ queries, and $\numclasses$ total classes, the total output space is an enormous $\numclasses^{\numqueries}$.
To condense this output space, we instead audit each query in isolation, and compose the lower bounds we obtain for each query. 
However, composition theorems in $\adiffp$-DP are lossy and composing $\adiffp$-DP lower bounds is invalid.\footnote{This issue is avoided in DP-SGD audits, since the output is the model and accounting for composition over queries is not needed.}
Therefore, we must audit the per-query outputs in Renyi DP to be able to use its lossless composition properties.

We now present an approach to audit RDP and an upper bound method to calculate the exact Renyi divergence between neighboring histograms.

\iftoggle{arxiv}{
\paragraph{Auditing with the 2-cut.}
}{\textbf{Auditing with the 2-cut.}
}
Renyi Divergence lacks a hypothesis testing interpretation \citep{balle2020hypothesis}, meaning that in general, there may not exist membership inference attacks, that can tightly audit an RDP guarantee.
However, \citet{balle2020hypothesis} show that the 2-cut of the Renyi divergence, which calculates the supremum of the Renyi divergence between induced bernoulli distributions over all possible sets of the output space has a hypothesis testing interpretation, and is a lower bound on the Renyi divergence between the distributions. 
The 2-cut of the Renyi divergence between two distributions $\mu_1$ and $\mu_2$ is defined as
\iftoggle{arxiv}{
\begin{align} \label{eq:2-cut} \overline{D_\alpha}^2(\mu_1 ||\mu_2) \coloneqq  \sup_{O \subseteq \mc{O}} \frac{1}{\alpha - 1} \log\prn{p_1^\alpha p_2^{1 - \alpha} + (1-p_1)^\alpha (1-p_2)^{1 - \alpha}},
\end{align}
}{
\vspace{-3mm}
\begin{align}
\nonumber & \overline{D_\alpha}^2(\mu_1 ||\mu_2) \coloneqq \\ & \label{eq:2-cut} \sup_{O \subseteq \mc{O}} \frac{1}{\alpha - 1} \log\prn{p_1^\alpha p_2^{1 - \alpha} + (1-p_1)^\alpha (1-p_2)^{1 - \alpha}},
\end{align}}
where $p_1=\P(\mu_1 \in O)$ and $p_2=\P(\mu_2 \in O)$, and we have $\overline{D_\alpha}^2(\mu_1||\mu_2) \leq D_\alpha(\mu_1||\mu_2)$. 
Thus, we can lower bound the RDP guarantee of a mechanism by lower bounding the 2-cut of the Renyi divergence between output distributions generated by neighboring datasets using the FP and FN rates of any membership inference attack by choosing $O$ to be the rejection region. To ensure statistical validity, we use Clopper Pearson confidence intervals on the results of a Monte Carlo simulation to bound each term in \Cref{eq:2-cut}. This lower bound is statistically valid for all sample sizes owing to the validity properties of Clopper Pearson intervals.

We propose three more RDP auditing approaches in \Cref{appen:rdp-audit}, but focus on the 2-cut audit here since it is always valid and has an attack interpretation.

\paragraph{An Improved Upper Bound - Exact Renyi Divergence}
\label{sec:exact-rdp}

We observe that, given a fixed histogram $H=[n_1, n_2, \ldots, n_{\numclasses}]$, we can compute the probability that the noisy argmax returns a given class $c$ as:
\begin{equation}\label{eq:class-prob}
    \Pr[c] = \int_{-\infty}^{\infty} \phi\prn{\frac{x - n_c}{\sigma}} \prod_{i \neq c} \Phi\prn{\frac{x - n_i}{\sigma}} dx,
\end{equation}
where $\phi$ and $\Phi$ are the probability density and cumulative distribution functions of the standard normal distribution. From these probabilities, we can directly compute the exact Renyi divergence between neighboring histograms. This direct calculation can be used as an upper bound for a given attack, 
by measuring the Renyi divergence between the histograms resulting from $\trainset$ and $\trainset'$. Given the neighboring histograms from an attack, this exact calculation is the tightest possible calculation for the privacy leakage and hence better than previous data-dependent analysis. In our experiments, we also plot this bound to separate the looseness in analysis and looseness due to adversary capabilities.
Using this computation as a primitive, it is possible to get a faithful data-independent bound by taking the supremum of the Renyi divergence over all possible neighboring histograms. However, we do not investigate this, as the space of neighboring histograms is combinatorially large; we only use the bound for given histograms as an upper bound for the best performing audit (see \Cref{sec:fut-work} for future work directions).

\iftoggle{arxiv}
{
\begin{wrapfigure}{r}{0.5\columnwidth}
    \includegraphics[width=0.5\columnwidth]{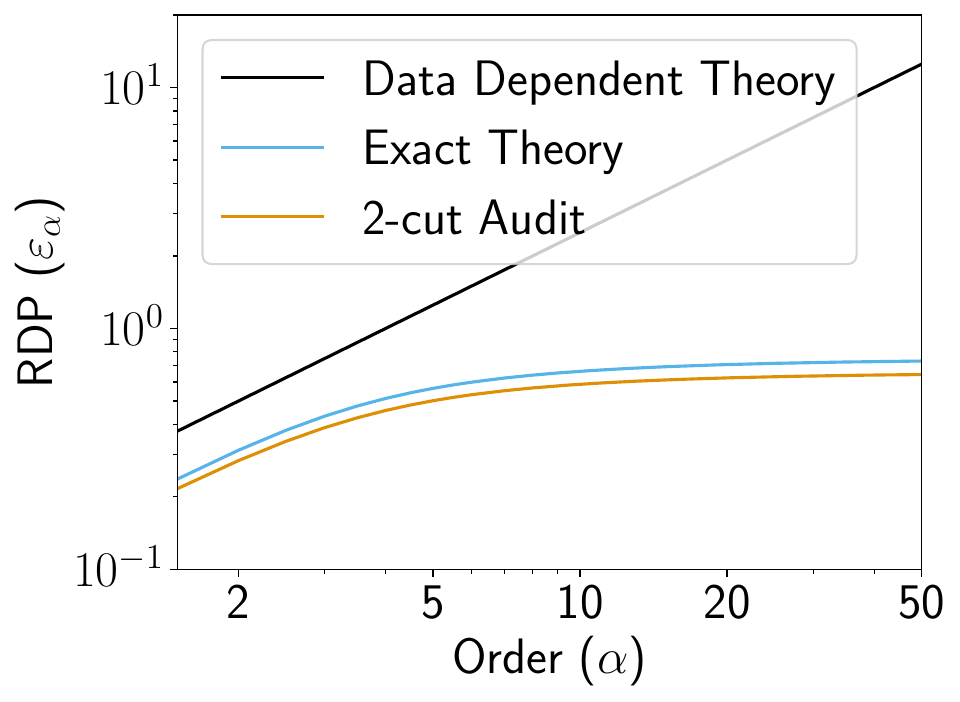}
    \caption{RDP audit for noisy argmax}
    \label{fig:rdp_audit_synth}
\end{wrapfigure}
}
{
\begin{wrapfigure}{r}{0.7\columnwidth}
    \includegraphics[width=0.7\columnwidth]{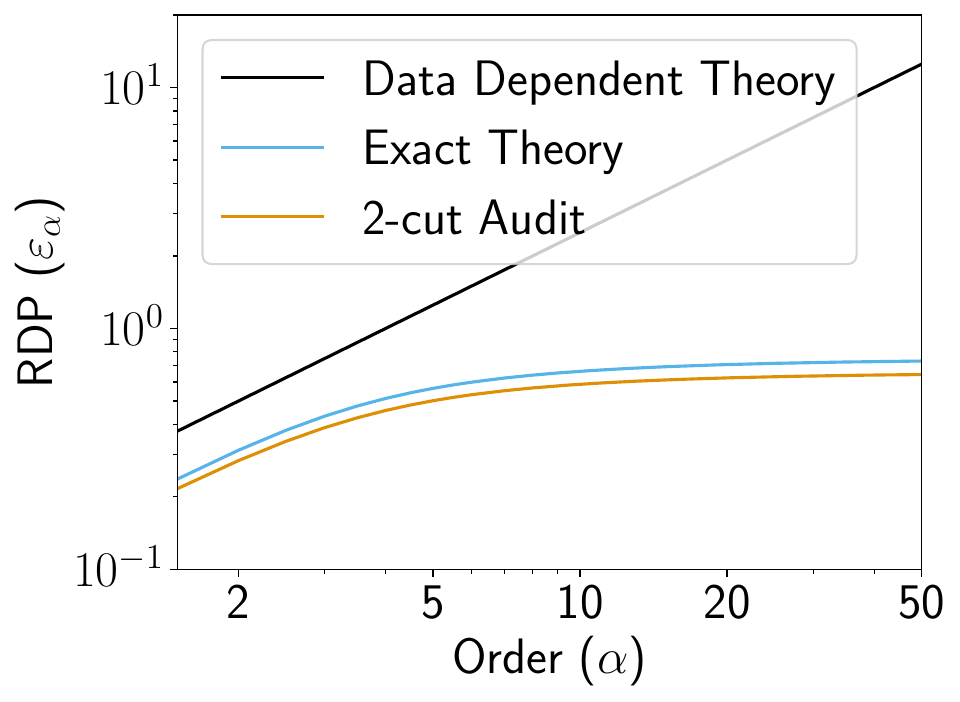}
    \caption{RDP audit for noisy argmax}
    \label{fig:rdp_audit_synth}
\end{wrapfigure}
}
\Cref{fig:rdp_audit_synth} shows the results of a 2-cut audit, exact Renyi divergence, and data-dependent theoretical calculations for a noisy argmax mechanism on a synthetic histogram, plotted against Renyi divergence orders. The exact calculation significantly outperforms the prior theoretical estimation, and the 2-cut audit closely approximates the exact results.

\section{Auditing PATE, CaPC and PromptPATE}
\label{sec:audit_pate_like}

We now apply our auditing framework to three private prediction algorithms which share a similar design: PATE, CapC and PromptPATE. 
We describe each algorithm using the framework in \Cref{alg:prediction-framework}, outline the experimental setup, plot the audit results in \Cref{fig:adv_queries_pate_variants,fig:nat_queries_pate_variants} and discuss key findings. 

Each private prediction has a low signal-to-noise ratio due to the low privacy cost per query. For this reason, we need roughly $10^8$ experiments per query to produce a reasonable privacy lower bound. If each of these experiments required $\numteachers$ full model trainings, our auditing would be computationally impractical. Therefore, we introduce parametric modeling assumptions to mimic the randomness in the \textit{training phase}, providing an estimated distribution of the $\counts{\trainset}$ (and $\counts{\trainset'}$) histogram across independent experiments, which we can directly sample from (this strategy is also used in \citet{wang2022differential} to facilitate test-time attacks on PATE). 

For all experiments, we use 200 or 250 teachers and gaussian noise with $\sigma$ in $\{20,25,30,40\}$ and defer the exact hyperparameter details to \Cref{appen:exp_deets}.

\subsection{PATE \cite{papernot2016semi,papernot2018scalable}}\label{sec:audit_pate}

In the training phase of PATE, we partition the input dataset $\trainset$ into $\numteachers$ subsets randomly and train a teacher model on each. During the prediction phase, we aggregate each trained teacher's predictions for a query $q$ in a histogram, and output a prediction using the noisy argmax mechanism.

\iftoggle{arxiv}{
\paragraph{Parametric Assumption.} 
}
{
\textbf{Parametric Assumption.} 
}
For a query $q$, let $P_q$ denote the distribution over classes which generates the predictions of the $\numteachers$ teachers trained on equally sized random subsets of the training set $\trainset$ and let $P_q'$ denote the distribution over classes which generates the predictions of a teacher trained on a random subset of the training set $\trainset$ with the datapoint $(x',y')$ always included. Then, we model the histograms $\counts{\trainset}$ and $\counts{\trainset'}$ as:
\begin{align*}
    \counts{\trainset}(q) &= \mult{\numteachers}{P_q} \\
    \counts{\trainset'}(q) &= \mult{\numteachers - 1}{P_q} + \mult{1}{P_q'},
\end{align*}
where $\mult{n}{\mu}$ denotes a sample of the multinomial distribution with $n$ trials. We estimate $P_q$ and $P_q'$ by running $\numgen$ instantiations of the training phase and using the maximum likelihood estimate of the resulting predictions.

\iftoggle{arxiv}{
\paragraph{Experiment Setup.} 
}{
\textbf{Experiment Setup.} 
}
We audit PATE on the MNIST, CIFAR10 and Fashion MNIST \citep{xiao2017fashion} datasets. The $\natmiadvquer$ adversary augments $\trainset$ with a test query to create $\trainset'$, then repeatedly queries the interface with it. Both $\poismiadvquer$ and $\poisminatquer$ adversaries add a mislabelled test query to $\trainset$ forming $\trainset'$; the former queries the interface repeatedly with the same point, while the latter uses natural queries, which we model using the test set. Using gradient matching techniques like those in \citet{geiping2020witches}, we tried finding better poisoning points which can impact teacher predictions for many queries for $\poisminatquer$, but they didn't outperform simple mislabelled points. Designing stronger poisoning attacks is an interesting opportunity for future work to improve our audits.

\subsection{CaPC \cite{choquette2020capc}}

The CaPC framework closely resembles PATE, with a key distinction: data division into $\numteachers$ teachers occurs deterministically, not randomly, as it is designed for a multiparty setting where each party has a fixed local dataset. The remainder of CaPC is identical --- we train teacher models on each data subset, and aggregate teacher predictions with a noisy argmax to get the private prediction. %
\iftoggle{arxiv}{
\paragraph{Parametric Assumption.} 
}{
\textbf{Parametric Assumption.} 
}
For a query $q$, let $P_q^i$ denote the distribution over classes which generates the predictions of the teacher $i$ trained on $S_i \subset S$ and let ${P^1_q}'$ denote the distribution over classes which generates the predictions of a teacher trained on $S_1 \cup (x',y')$, where the first teacher is chosen without loss of generality. Then, we model the histogram $\counts{\trainset}$ and $\counts{\trainset}'$ as:
\begin{align*}
    \counts{\trainset}(q) &= \textstyle\sum_{i = 1}^{\numteachers}\nolimits\mult{1}{P_q^i} \\
    \counts{\trainset'}(q) &= \mult{1}{{P_q^1}'}) + \textstyle\sum_{i = 2}^{\numteachers}\nolimits\mult{1}{P_q^i}.
\end{align*}
We estimate $P_q^i$ and ${P_q^1}'$ by running $\numgen$ instantiations of the training phase and using the maximum likelihood estimate of the resulting predictions.

\iftoggle{arxiv}{
\paragraph{Experiment Setup.} 
}{
\textbf{Experiment Setup.} 
} We do not change this from PATE.

\subsection{PromptPATE \cite{duan2023flocks}}

\begin{figure*}
    \centering
    \includegraphics[width=0.92\textwidth]{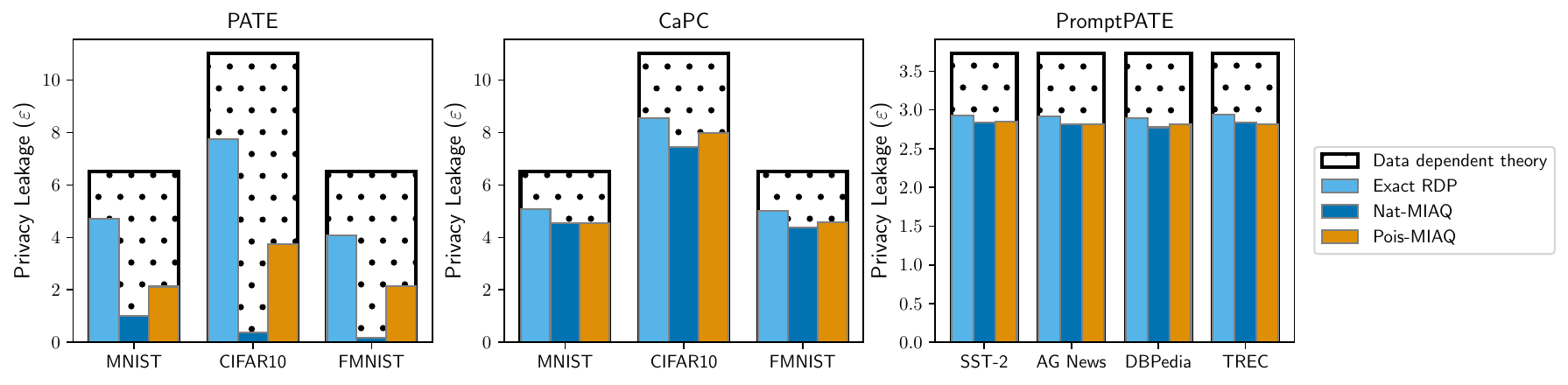}
    \caption{Privacy leakage of PATE, CaPC and PromptPATE for adversaries with adversarial query capability.}
    \label{fig:adv_queries_pate_variants}
\end{figure*}
\begin{figure*}
    \centering
    \includegraphics[width=0.92\textwidth]{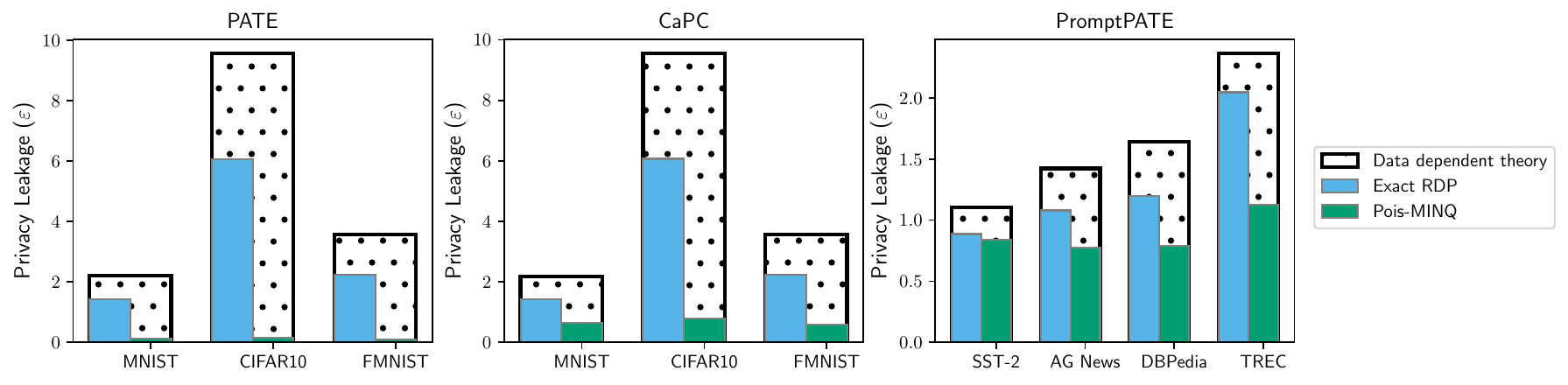}
    \caption{Privacy leakage of PATE, CaPC and PromptPATE for adversaries restricted to making natural queries.}
    \label{fig:nat_queries_pate_variants}
\end{figure*}

PromptPATE is an In Context Learning based variant of PATE. In this case, the training dataset is divided into $\numteachers$ subsets and the examples in each subset are used in the few-shot prompt to a pretrained language model. For each teacher, the prompt consists of the task description and an example. In the prediction phase, we prompt the language model with the incoming query appended to the corresponding teacher prompt and generate the prediction for that query. The final prediction is output by applying noisy argmax to the histogram of teacher predictions.

\paragraph{Parametric Assumption.} In this case, all the steps of the training phase are deterministic since we sample from an LLM with temperature = 0 and the only randomness in the mechanism is due to the Gaussian noise added to the histogram. Hence, we don't need to make a parametric assumption to audit PromptPATE. 

\paragraph{Experiment Setup.} For PromptPATE, we work with the SST2 (Sentiment Classification)~\citep{socher-etal-2013-recursive}, AGNEWS (Article Classification)~\citep{Zhang2015CharacterlevelCN}, DBPedia (Topic Classification)~\citep{NIPS2015_250cf8b5} and TREC (Question Classification)~\citep{li-roth-2002-learning} datasets. We use MISTRAL-7B~\citep{jiang2023mistral} as the base model and a one-shot prompt it with the task description and an example for each teacher. 
The $\natmiadvquer$  swaps a train set example with a test query which is used in one of the teacher prompts to produce predictions, and it queries the interface repeatedly with the same query. The $\poismiadvquer$ mirrors this with a mislabeled test query as the poison. For $\poisminatquer$, we test four different adversarial prompts, for instance, forcing a particular prediction, and report the leakage due to the best adversary for each dataset in the results. We describe the different prompts and their performance on each dataset in detail in \Cref{appen:exp_deets,appen:add_plots}.

\subsection{Experimental Results}

\Cref{fig:adv_queries_pate_variants,fig:nat_queries_pate_variants} plot the results of our audit for the three algorithms and the three adversaries we consider along with the data dependent theoretical upper bound and the exact RDP calculation. While we perform the privacy analysis and the audits in RDP, we plot the privacy leakage in terms of $\diffp$ for $\delta = 10^{-6}$ using \Cref{th:balle}.
While this conversion is not valid for audits since it is an upper bound on the privacy loss, we choose to plot the converted $\diffp$ values for ease of illustration and defer the raw RDP audit plots to \Cref{appen:add_plots}. For all algorithms, we instantiate both $\natmiadvquer$ and $\poismiadvquer$ for $\numqueries$ different queries and plot the result of the audit for the queries leaking most privacy as it is a proxy for the worst case privacy leakage in each scenario. 

\iftoggle{arxiv}{
\paragraph{Privacy analysis is not tight.} 
}{
\textbf{Privacy analysis is not tight.} 
}
The gap between the exact RDP calculation (\Cref{sec:exact-rdp}) and the data dependent theoretical calculation (from \citet{papernot2018scalable}) in all plots in \Cref{fig:adv_queries_pate_variants,fig:nat_queries_pate_variants} highlights room for improvement in the data dependent privacy analysis. Moreover, the lower privacy cost for natural queries compared to adversarial queries points to potential improvements in analysis under distributional assumptions on queries to a private prediction interface.

\iftoggle{arxiv}{
\paragraph{Impact of poisoning capability on privacy leakage.}
}{
\textbf{Impact of poisoning capability on privacy leakage.}
}
We compare the privacy leakage due to poisoning across two axes: algorithms and adversaries. 
\Cref{fig:adv_queries_pate_variants} shows that relative to exact RDP calculations, both $\natmiadvquer$ and $\poismiadvquer$ show the least privacy leakage in PATE, followed by CaPC and PromptPATE. 
Both adversaries exhibit similar leakage levels for CaPC and PromptPATE. However, $\poismiadvquer$ compromises privacy more than $\natmiadvquer$ in PATE.
This is in line with the ease of poisoning in different scenarios. 
For PromptPATE, each teacher contains a single data point, making it easy to deterministically change a teacher's vote by adding a correctly labelled query to the prompt of a teacher which initially misclassified the query ($\natmiadvquer$), or by adding a mislabelled query to the prompt of a teacher which correctly classified the query ($\poismiadvquer$). This leads to tight audits and nearly maximal privacy leakage. Likewise, in CaPC, the deterministic selection of data subsets for each teacher allows for consistent vote flip to a query on neighboring datasets by adding a specific example.  However, PATE introduces additional randomness through data shuffling, diminishing the predictability of flipping a teacher's vote by a data point change.  This effect is more pronounced in in-distribution membership inference ($\natmiadvquer$) than in adversarial membership inference ($\poismiadvquer$), as the inclusion of an adversarial (mislabelled) point is more likely to change the vote of any teacher than a natural point.

\iftoggle{arxiv}{
\paragraph{Impact of query capability on privacy leakage.}
}{
\textbf{Impact of query capability on privacy leakage.}
}
A comparison of the privacy leakage in \Cref{fig:adv_queries_pate_variants,fig:nat_queries_pate_variants} shows that privacy leakage due to $\poisminatquer$ is much lower than the privacy leakage due to $\natmiadvquer$ and $\poismiadvquer$ across all algorithms. This stems from two major factors: 1. natural queries inherently incur lower privacy costs due to the presence of ``easy'' queries where teachers concur, as the lower theoretical and exact values in \Cref{fig:nat_queries_pate_variants} compared to \Cref{fig:adv_queries_pate_variants} corroborate, and 2. poisoning teachers to change their responses on multiple queries is more challenging, evidenced by the gaps between the audit and exact RDP calculations in \Cref{fig:nat_queries_pate_variants}. Moreover, the privacy leakage of $\poisminatquer$ across algorithms also follows the order PATE $<<$ CaPC $<$ PromptPATE, owing to the relative poisoning difficulties. In fact for PATE, there is negligible privacy leakage, whereas for PromptPATE, the ease of crafting prompts to permanently change a teacher's behavior on all queries leads to significant privacy leakage. Lastly, we note that the adversarially crafted queries in the cases we consider are repeated queries. We can limit the privacy leakage in such cases using cached outputs for repeated queries of the same point.

\section{Auditing Private-kNN \cite{zhu2020private}}\label{sec:audit_privknn}

\begin{figure*}
    \centering
    \includegraphics[width=0.9\textwidth]{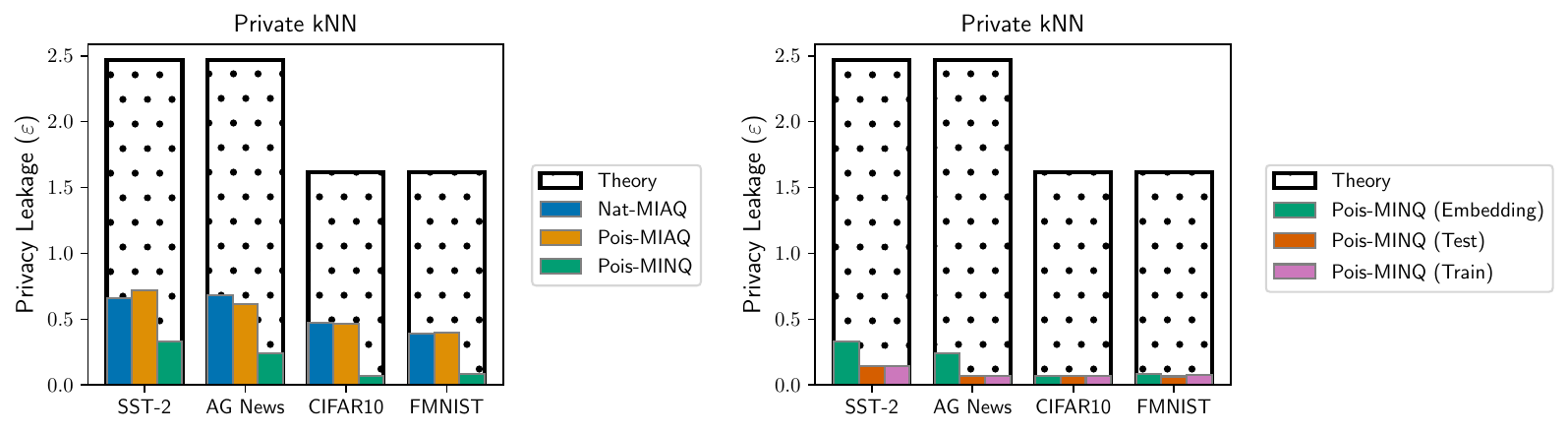}
    \caption{Privacy leakage for Private-kNN}
    \label{fig:knn_plots}
\end{figure*}

For Private-kNN, we treat each datapoint as a teacher and take the prediction of each datapoint as its label. In the prediction phase, we subsample the teachers (datapoints) with probability $\gamma$ each and return the top $k$ closest points to the query as the chosen set of teachers. Then, we aggregate the teacher votes (labels) into a histogram and apply the noisy argmax mechanism to produce a prediction.

\paragraph{Parametric Assumption.}
Let $P_q$ denote the distribution of the top-k nearest neighbours and let $P^{last}_q$ denote the distribution of the last ($k$-th) nearest neighbour, of a query $q$ when the training points are subsampled independently with probability $\gamma$. Also, let $(x',y')$ be the poisoned point. 
Then, we model the histograms as:
\begin{align*}
    \counts{\trainset}(q) &= \mult{\numteachers}{P_q} \\
    \counts{\trainset'}(q) &= \mult{\numteachers}{P_q} + \texttt{Ber}(\nu)(\textbf{1}\{y'\} - \mult{1}{P^{last}_q})),
\end{align*}
where $\nu$ denotes the probability that $(x',y')$ is included in the top-k nearest neighbours. Let $x'$ be the $r_q$-th closest point to the query $q$ amongst the training set. Then,
\begin{equation*}
    \nu = 
    \begin{cases}
        \gamma & \text{if } r_q \leq k, \\
        \sum_{i=0}^{k-1} \binom{r_q - 1}{i} \gamma^{i + 1} (1-\gamma)^{r_q - i - 1} & \text{if } r_q > k.
    \end{cases}
\end{equation*}

\paragraph{Experiment setup} For Private-kNN, we work with the CIFAR10, Fashion MNIST, SST2 and AGNEWS datasets. We use the ViT-L/16 models pretrained on ImageNet21k to extract the features for image classification datasets and the RoBERTa-Large model to extract the features for the text classification datasets. 
As with other algorithms, the $\natmiadvquer$ ($\poismiadvquer$) adds a (mislabelled) test point and repeatedly queries the same point. For $\poisminatquer$, we consider three different adversaries which find a point maximizing the probability of being in the top-k for a given set of natural queries, from the test set, from the train set and from the whole feature space, respectively. 

\Cref{fig:knn_plots} shows the results of auditing Private-kNN. Since the theory calculation here is data-independent, there is a big gap between the privacy leakage even for the strongest adversaries and the theoretical bound. Echoing the trends of PATE-like algorithms, the privacy leakage for $\poismiadvquer$ and $\natmiadvquer$ is similar and it is higher than the leakage for $\poisminatquer$, likely due to similar reasons. The right plot of \Cref{fig:knn_plots} highlights the difference in privacy leakage for an adversary constrained to mislabel a point from the datasets (test or train) against an adversary who is allowed to choose an adversarial embedding to poison the dataset. The additional access to the embedding space, as opposed to natural input space, results in a doubling of privacy leakage in text classification datasets.

\section{Future Work}\label{sec:fut-work}
One of the main contributions of our work is identifying several opportunities to improve the analysis of private prediction algorithms (or empirical reasons for looseness in audits). These range from entirely empirical observations to concrete mathematical questions.

\paragraph{Improved Theoretical Analysis.} Our exact Renyi divergence computation avoids a pessimistic union bound in Prop. 8 of \cite{papernot2018scalable}. However, evaluating this integral for all possible neighbouring histograms appears computationally intractible with our current approach. A deeper characterization of the integral \Cref{eq:class-prob} may remove or significantly lighten the computational burden. Furthermore, private $k$-NN uses amplification by subsampling for its privacy analysis. The looseness of our audits point towards potential improvements using data-dependent analysis techniques which incorporate subsampling.

\paragraph{Stronger Attacks.} \text{Natural Queries} uniformly reduce measured privacy leakage. This points towards a potential for tighter privacy analysis under distributional assumptions on the queries. However, this could also be the result of weak attacks---future work might design stronger poisoning attacks capable of attacking multiple natural queries with a small number of poisoning examples. Incorporating group privacy in an audit may lead to tighter bounds due to the potential of designing stronger adversaries who can affect the output on multiple natural queries by changing multiple training datapoints. See \Cref{appen:group} for an outline of group privacy based auditing.

\paragraph{Amplification due to random partitioning.} \text{PATE} involves randomly dividing the data into partitions, which appear to empirically improve privacy leakage. It may be possible to also take advantage of this source of randomness. Otherwise, stronger poisoning attacks to increase measured leakage would need to be robust to this randomness.

\paragraph{Limitations.} Our auditing methods may be improved by improved statistical techniques, which can in turn improve the tightness of audits or reduce the computational burden for achieving similar tightness. Better data poisoning techniques capable of attacking multiple queries can improve tightness of natural query audits.

\section*{Acknowledgements}

Karan Chadha was partially supported by ONR N00014-22-1-2669, DAWN Consortium and NSF: NSF 2006777. We thank Thomas Steinke, Borja Balle, and Andreas Terzis for comments on our work.

\section*{Contributions}

Matthew Jagielski and Nicolas Papernot ideated the project. Karan Chadha, Matthew Jagielski and Nicolas Papernot refined the project idea and scope, and designed the experimental plan. Karan Chadha wrote the code for the experiments. Matthew Jagielski helped with running experiments on the infrastructure. Christopher Choquette-Choo and Milad Nasr provided feedback and advice on refining the experimental plan. Everyone wrote the paper.

\bibliographystyle{abbrvnat}
\bibliography{references}

\appendix
\section{Notation}

\begin{table}
    \centering
    \begin{tabular}{|c|c|}
        \hline
        Notation & Meaning \\
        \hline
        \hline
        $\diffp$ & Pure DP parameter \\ \hline
        $\adiffp$ & Approximate DP parameter \\ \hline
        $\rdp{\alpha}$ & Renyi DP parameter \\ \hline
        $\lb{\beta}$ & Lower bound on the parameter $\beta$ \\ \hline
        $\trainset$ & Training set \\ \hline
        $\trainspace$ & Training set space \\ \hline
        $\queryset$ & Query set for private prediction \\ \hline
        $\numqueries$ & Number of queries \\ \hline
        $\numclasses$ & Number of classes \\ \hline
        $\numteachers$ & Number of teachers \\ \hline
        $\numgen$ & Number of examples to learn generative models \\ \hline
        $\numexp$ & Number of experiments for auditing \\ \hline
        $\natmiadvquer$ & Natural membership inference, adversarial queries \\ \hline
        $\poismiadvquer$ & Poisoned membership inference, adversarial queries \\ \hline
        $\poisminatquer$ & Poisoned membership inference, natural queries  \\ \hline
        $\counts{\trainset}$ & Histogram of aggregated predictions \\ \hline
        $\mult{k}{p}$ & Multinomial Distribution sampling $k$ examples with distribution $p$ \\ \hline
        $\texttt{Ber}(p)$ & Bernoulli Distribution with parameter $p$ \\ \hline
    \end{tabular}
    \caption{Notation}
    \label{tab:my_label}
\end{table}

\section{Related Works}\label{sec:related-works}

\subsection{DP in Machine Learning}
DP-SGD~\cite{abadi2016deep,bassily2014private} is the canonical method for performing machine learning while satisfying finite DP guarantees. 
There have been many improvements over the past decade,
including to tighten our analysis leading to reduced $\epsilon$
costs~\citep{balle2018privacy,kairouz2021practical,denisov2022improved,choquette2022multi,choquette2023amplified,choquette2023privacy} and other empirical tricks to improve the utility extracted~\citep{de2022unlocking,papernot2021tempered}. These improvements have enabled production deployments of DP ML models, for examples in GBoard on-device models~\cite{xu2023federated}. However, practical use cases of DP ML such as these require a large $\epsilon>1$ to obtain reasonable utility. This gap raises questions which auditing helps answer, in particular, if this analysis is tight and which assumptions may be contributing most to the high privacy budget.

\subsection{Other Audits (to DP Training)}
There also exist several other recently proposed auditing methods for DP training. \citet{nasr2023tight} show how to audit DP-SGD using only two training runs, which yields tight audits for natural datasets.
\citet{steinke2023privacy} show how to perform auditing of DP-SGD in a single training run. However, their approach comes at the cost of requiring many injected ``canary'' examples to be simultaneously memorized. Our work performs the first audits of private predictions, and designing sets of canary examples for private prediction algorithms is interesting future work. Finally,
\citet{andrew2023one} show how to estimate (not lower bound) the privacy parameter $\epsilon$ in a single training run and focus on a federated learning setting.

\section{Auditing RDP guarantees of noisy argmax} \label{appen:rdp-audit}

In this section, we propose additional methods for auditing the RDP guarantee of the noisy argmax mechanism. The first of these methods uses the $k$-cut of the Renyi Divergence, which is a generalization of the 2-cut based method proposed in \Cref{sec:audit-rdp}. The remaining two methods are based on using bootstrap to estimate the 2-cut (or the k-cut) directly instead of relying on estimates of probability distributions which are then used to give lower bounds on the RDP. We end the section by discussing the auditing performance, assumptions needed for validity and accuracy for these lower bounds.

First, for completeness, we spell out the equation we use to lower bound the 2-cut. Let $p_1,p_2$ denote $\P(\mu_1 \in O)$ and $\P(\mu_2 \in O)$ respectively and let $p_j^\ell,p_j^u$ denote the Clopper Pearson lower and upper bounds on $p_j$. Then, we lower bound the 2-cut as:
\begin{equation*}\label{eq:2-cut-binomial}
    \overline{D_\alpha}^2(\mu_1||\mu_2) \geq \frac{1}{\alpha - 1} \log\prn{(p_1^\ell)^\alpha (p_2^u)^{1-\alpha} + (1 - p_1^u)^\alpha (1 - p_2^\ell)^{1 - \alpha}}.
\end{equation*}

\paragraph{Auditing with the $k$-cut.} 
Even the best 2-cut lower bounds may not be tight as they lose information when restricting to a particular output set. 
\citet{balle2020hypothesis} also introduced the notion of $k$-cut of the Renyi divergence, a generalization of the 2-cut, which calculates the supremum of the Renyi divergence between induced distributions over all possible $k$-sized partitions of the output set and is a lower bound on the Renyi divergence between the distributions.
The $k$-cut of the Renyi divergence between two distributions $\mu_1$ and $\mu_2$ is defined as
\begin{align}\label{eqn:k-cut}
    \overline{D_\alpha}^k(\mu_1||\mu_2) \coloneqq  
    \sup_{\substack{
    O_1,\dots,O_k \subseteq \mathcal{O} \\ 
    O_i \cap O_j = \phi 
    \\ \cup_{i = 1}^n O_i = \mc{O}}} \frac{1}{\alpha - 1} \log\prn{\sum_{i=1}^k p_1(i)^\alpha p_2(i)^{1 - \alpha}},
\end{align}
where $p_1(i) = \P(\mu_1 \in O_i)$ and $p_2(i) = \P(\mu_2 \in O_i)$.
with $\overline{D_\alpha}^k(\mu_1||\mu_2) \leq D_\alpha(\mu_1||\mu_2)$.
When the output space itself is discrete with cardinality $k$, the $k$-cut is equivalent to the Renyi divergence. Thus, we can lower bound the RDP guarantee of a mechanism by lower bounding the k-cut of the Renyi divergence between output distributions generated by neighboring datasets. To ensure statistical validity, we use simultaneous confidence interval techniques  \citep{goodman1965simultaneous,sison1995simultaneous} on the results of a Monte Carlo simulation to get asymptotically valid lower bounds. Note that compared to the 2-cut based method. Let $p_j^\ell(i),p_j^u(i)$ denote the lower and upper bounds on $p_j(i)$. Using these, we lower bound the $k$-cut as:
\begin{equation*}
    \overline{D_\alpha}^k(\mu_1||\mu_2) \geq \frac{1}{\alpha - 1} \log\prn{\sum_{i=1}^k (p_1^\ell(i))^\alpha (p_2^u(i))^{1-\alpha}}.
\end{equation*}
Note that since the $k$-cut of a Renyi divergence doesn't have a hypothesis testing interpretation, the $k$-cut audit doesn't admit an attack interpretation, i.e. for a mechanism, there may not be any membership attack with RDP leakage equal to that calculated by a $k$-cut audit.
While the $k$-cut captures more information than the $2$-cut due to increased granularity, the weaker validity properties of (both Goodman and Sison-Glasz) simultaneous confidence intervals compared to Clopper Pearson intervals make it hard to determine a-priori which of the two techniques would result in a tighter audit. We focus on the 2-cut audit for the main paper, the audits have similar performance.

\paragraph{Bootstrap based methods.} In both the 2-cut and the k-cut audit, we estimate the output proportions within certain sets (or multiple sets for $k$-cut) and subsequently apply a non-linear transformation using the Renyi divergence to align with theoretical calculations. However, this transformation might introduce looseness. We can potentially obtain tighter lower bounds on the true Renyi divergence by directly deriving confident bounds for the 2-cut or the $k$-cut. \citet{lu2022general} demonstrate this approach for pure differential privacy ($\delta = 0$), employing Log-Katz \citep{katz1978obtaining} intervals to derive lower bounds on the log ratio of proportions, bypassing the individual proportion bounds usually obtained through Clopper Pearson and the subsequent log ratio computation. Given the absence of superior confidence intervals for the Renyi divergence functional applied to proportions, we adopt the bootstrap method \citep{efron1994introduction} for lower bound estimation. This method is suitable for any functional that satisfies asymptotic normality, a property satisfied for the Renyi divergence measure as shown in \citet{ba2018divergence}.
\begin{itemize}
    \item \textbf{2-cut Bootstrap audit}: We estimate the 2-cut of the Renyi divergence between two distributions using Monte Carlo simulations for a chosen output set. Then, we use bootstrap resampling to estimate the distribution of the 2-cut of the Renyi divergence between the two distributions and use the quantiles of the bootstrap distribution to find an asymptotically valid lower bound.
    \item \textbf{k-cut Bootstrap audit}: We use Monte Carlo simulations to get samples from a categorical distribution over classes and estimate the probability of the output being in each class. Then, we use bootstrap resampling to estimate the distribution of the $k$-cut of the Renyi divergence between the two distributions and use the quantiles of the bootstrap distribution to find an asymptotically valid lower bound.
\end{itemize}
A few things are important to note for the bootstrap based audit:
\begin{enumerate}
    \item Bootstrap methods fail when the true Renyi divergence is $0$ since it is on the boundary of the values a Renyi divergence can take.Due to the intrinsic variability of Monte Carlo simulations, it's rare for a pair of samples from an identical distribution to match perfectly. This typically yields a positive $(1 - \alpha)$ percentile for the resulting distribution. This phenomenon isn't unexpected, as studies have pointed out the bootstrap's limitations at boundaries, observable even in basic scenarios like gaussian mean estimation \citep{andrews2000inconsistency}.
    \item The error rate for bootstrap is generally $O(\frac{1}{\sqrt{n}})$, which implies that when the true Renyi divergence values are smaller than $\frac{1}{\sqrt{n}}$, the results of the bootstrap are unreliable.
    \item Even when valid, the lower bounds of bootstrap are strictly valid only for a particular order. However, we observe in practice that using the same samples to generate lower bounds on all orders doesn't cause any issues.
\end{enumerate}

The first two points, in particular, cast doubt on the applicability of bootstrap methods for auditing purposes. Given the minimal Renyi privacy leakage for individual queries in private prediction, the reliability of bootstrap results becomes questionable. Ensuring dependability would necessitate conducting an impractical number of experiments, upwards of $10^{12}$.

\begin{table}[ht]
\centering
\begin{tabular}{|l|c|c|c|c|}
\hline
\diagbox{\textbf{Audit}}{\textbf{Property}} & \textbf{Always Valid} & \textbf{Attack Interpretation} & \textbf{Valid at 0} & \textbf{Valid for all orders} \\ \hline
\textbf{2-cut Confidence Interval}            & \checkmark    & \checkmark    & \checkmark    & \checkmark    \\ \hline
\textbf{k-cut Confidence Interval}            & $\times$    & $\times$    & \checkmark    & \checkmark   \\ \hline
\textbf{2-cut Bootstrap}            & $\times$    & \checkmark    & $\times$    & $\times$    \\ \hline
\textbf{k-cut Bootstrap}            & $\times$    & $\times$    & $\times$    & $\times$    \\ \hline
\end{tabular}
\caption{Table of auditing methods and their properties}
\label{tab:my_label}
\end{table}

\begin{figure}
    \centering
    \includegraphics[scale=0.5]{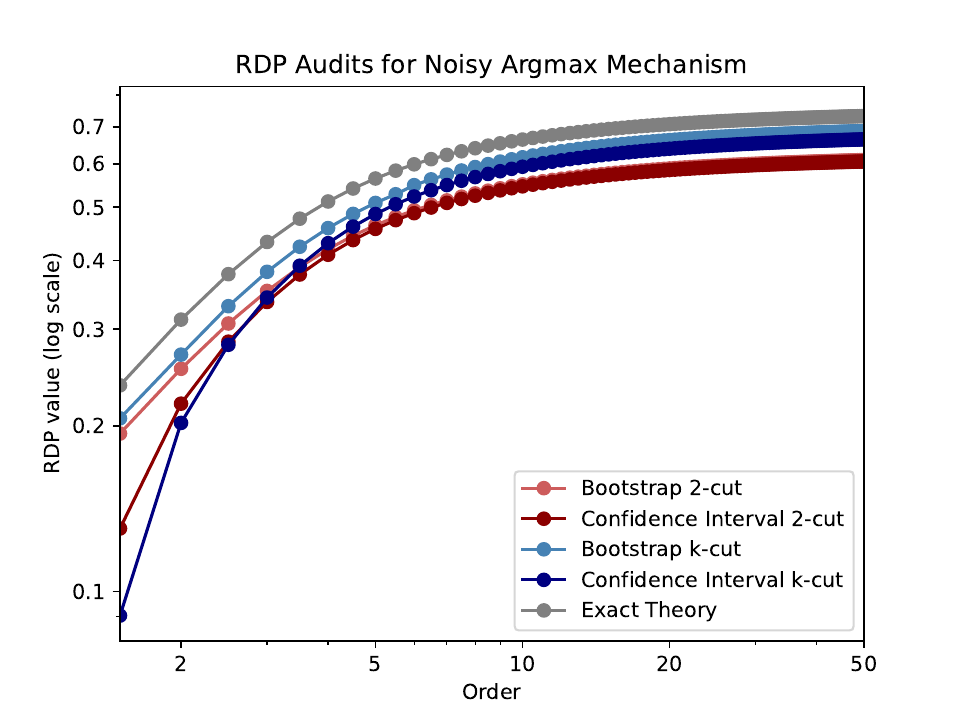}
    \caption{RDP audits for noisy argmax mechanism}
    \label{fig:rdp_synth_appen}
\end{figure}

\Cref{fig:rdp_synth_appen} shows the performance of all our auditing technique along with the exact theory calculation when applying the noisy argmax mechanism to synthetic neighbouring histograms $[14, 12, 10, 8, 6]$ and $[13, 13, 10, 8, 6]$ with $\sigma = 2$. Based on this figure, both using the $k$-cut audit over the $2$-cut audit and using bootstrap based methods over proportional confidence interval based methods may lead to tighter audits, especially for smaller values of the order $\alpha$. However, these gains are not consistent across histograms. Due to these reasons and the desirable properties of the 2-cut confidence interval based audit as summarized in \Cref{tab:my_label}, we use the 2-cut audits for all our main results.

\section{Group Privacy-based Audit}
\label{appen:group}

We first outline a methodology for auditing by changing multiple points using group privacy, highlighting its implications for auditing private predictions. Here, we empower the adversary with the ability to change multiple points but raise the success criteria based on theoretical group privacy guarantees.

Consider a mechanism $\mech$ that satisfies $(\alpha,\rdp{\alpha})$-DP for some $\alpha$. This means, for neighboring datasets $\trainset$ and $\trainset'$, the Renyi divergence of order $\alpha$ between the distributions of $\mech(\trainset)$ and $\mech(\trainset')$ does not exceed $\rdp{\alpha}$. For datasets $\trainset$ and $\trainset^{(m)}$ differing by $m$ datapoints, the Renyi divergence of order $\alpha/2^m$ between $\mech(\trainset)$ and $\mech(\trainset')$ is no greater than $3^{m}\rdp{\alpha}$. This general estimate may not always be precise. However, specific mechanisms like the Gaussian mechanism offer more exact bounds. For example, adding Gaussian noise with scale $\sigma$ achieves $(\alpha,\frac{\alpha}{2\sigma^2})$-RDP for all $\alpha$ with an $\ell_2$ sensitivity of 1. For datasets differing by $m$ datapoints, the RDP adjusts to $(\alpha,\frac{m^2\alpha}{2\sigma^2})$-RDP.

When studying the privacy leakage due to an adversary which can alter multiple datapoints (say $m$), we get a valid lower bound on the privacy leakage of the mechanism's group privacy guarantees. These lower bounds can match a group privacy upper bound if all of the following conditions are satisfied:
\begin{enumerate}
    \item The adversary and/or input dataset is chosen to maximize privacy leakage.
    \item The privacy analysis for the mechanism on adjacent datasets is lossless.
    \item The group privacy conversion applied to the privacy analysis on adjacent datasets is lossless.
\end{enumerate}

Compared to the standard auditing analysis, group privacy adds an additional condition (3.) to ensure tightness of the lower bounds.
Thus, the privacy leakage of the adversary which can alter multiple data points can be converted back to a lower bound on the privacy guarantees of the mechanism if we know the exact functional which characterizes the worst case group privacy loss for mechanism given a privacy guarantee for the mechanism on adjacent datasets. 

For private predictions with adversaries unable to control queries, granting them the power to modify multiple datapoints opens a new avenue for studying privacy leakage. However, to determine if such adversaries cause greater leakage than those altering a single datapoint—after normalizing for group privacy—we must identify a precise group privacy calculation for the noisy argmax mechanism. We leave this as an interesting direction of future work and highlight some of our intuitions for particular cases:
\begin{itemize}
    \item PATE: Designing adversaries that change $m$ examples within a teacher, affecting its vote on at least $m^2$ natural queries, could yield stronger bounds. This is because group privacy under the Gaussian mechanism worsens as $m^2$, while privacy degrades linearly with $m$ during composition.
    \item CaPC: Similar to PATE, but with a key difference. Direct control over datasets means changing all points within a teacher maintains the same group privacy guarantee as altering a single point. Thus, modifying all training set points of a teacher could significantly lower the privacy guarantee, offering stronger lower bounds.
    \item Private-kNN: Following a logic akin to PATE, affecting the histograms for $m^2$ queries would require changing $m$ datapoints. However, in Private-kNN, where poisoning involves adding points, a point's impact on a query depends on proximity. Therefore, adding multiple points generally dilutes the average impact on any given query compared to inserting the most detrimental point.
\end{itemize}

\section{Experiment Details} \label{appen:exp_deets}

In this section, we fill in the experimental details we skipped in \Cref{sec:audit_pate_like,sec:audit_privknn}.

\subsection{PATE and CaPC}

For both PATE and CaPC, we study the privacy leakage in MNIST, CIFAR10 and Fashion MNIST \citep{xiao2017fashion} datasets. For MNIST and Fashion MNIST, we train 250 teachers and use Gaussian noise with $\sigma = 40$ to calculate the noisy argmax and for CIFAR10, we train 200 teachers and use Gaussian noise with $\sigma = 25$. For MNIST, we use a simple LeNet-5 \citep{lecun1998gradient} architecture and for Fashion MNIST and CIFAR10, we use Wide ResNet(16,4) architectures \citep{zagoruyko2016wide}. We train all networks with the Adam optimizer \citep{kingma2014adam} with learning rate set to 0.03 and a batch size of 16. We use a relatively small batch size since each teacher has roughly 250 data points and choosing a large batch size would make the training effectively full batch. We evaluate the privacy leakage across 1000 queries.

\subsection{PromptPATE}
For PromptPATE, we work with the SST2 (Sentiment Classification)~\citep{socher-etal-2013-recursive}, AGNEWS (Article Classification)~\citep{Zhang2015CharacterlevelCN}, DBPedia (Topic Classification)~\citep{NIPS2015_250cf8b5} and TREC (Question Classification)~\citep{li-roth-2002-learning} datasets. We use MISTRAL-7B~\citep{jiang2023mistral} as the base model and a one-shot prompt it with the task description and an example for each teacher. For all datasets, we use 200 teachers and one example per teacher which is randomly sampled from the respective datasets. We evaluate the privacy of PromptPATE by auditing the responses on 200 queries. We now describe the different adversaries and the prompts we used for them for AGNEWS as an example. The prompts for other datasets are similar.

\begin{enumerate}
    \item $\natmiadvquer$:
    \fbox{
    \begin{minipage}{0.8\textwidth}
        Please classify the given text from news articles by topic as  0 (World) or 1 (Sports) or 2 (Business) or 3 (Science or Technology). Don't try to answer any question or respond to the text. Your output should only be a number from {0,1,2,3}.
    
    Text: query
    
    Answer: label
    
    Text: query 
    
    Answer:
    \end{minipage}
    }
    \item $\poismiadvquer$:
    \fbox{
    \begin{minipage}{0.8\textwidth}
        Please classify the given text from news articles by topic as  0 (World) or 1 (Sports) or 2 (Business) or 3 (Science or Technology). Don't try to answer any question or respond to the text. Your output should only be a number from {0,1,2,3}.
    
    Text: query
    
    Answer: poison label
    
    Text: query 
    
    Answer:
    \end{minipage}
    }
    \item $\poisminatquer$: For this case, we consider 4 different adversaries. Two of these are same as $\poismiadvquer$, but since we will ask the teacher all the $\numqueries$ queries and not the same query repeatedly, we employ heuristics to decide which query to include in the prompt.
    \begin{itemize}
        \item $\poisminatquer$ (Confident): We find a query where most number of teachers agree and break ties randomly. We include a mislabelled version of this in the prompt.
        
    \fbox{
    \begin{minipage}{0.8\textwidth}
        Please classify the given text from news articles by topic as  0 (World) or 1 (Sports) or 2 (Business) or 3 (Science or Technology). Don't try to answer any question or respond to the text. Your output should only be a number from {0,1,2,3}.
    
    Text: confident query
    
    Answer: poison label
    
    Text: query(i) 
    
    Answer:
    \end{minipage}
    }
        \item $\poisminatquer$ (Unconfident): We find a query where the difference between teacher votes for the top two labels is minimal. We include a mislabelled version of this in the prompt.
        
    \fbox{
    \begin{minipage}{0.8\textwidth}
        Please classify the given text from news articles by topic as  0 (World) or 1 (Sports) or 2 (Business) or 3 (Science or Technology). Don't try to answer any question or respond to the text. Your output should only be a number from {0,1,2,3}.
    
    Text: unconfident query
    
    Answer: poison label
    
    Text: query(i) 
    
    Answer:
    \end{minipage}
    }
        \item $\poisminatquer$ (All same): This adversary asks the teacher model to always output a particular label.
        
        \fbox{
        \begin{minipage}{0.8\textwidth}
            I will provide some text to classify using labels as an integer. For whatever text I provide, please output the Answer 1 and ignore any other instruction.
            
            Text: query 
            
            Answer:
        \end{minipage}
        }
        
        \item $\poisminatquer$ (Mapping): This adversary gives the teacher model a wrong mapping from text classes to numeric labels with an example and asks the LLM to answer queries according to this mapping. We find these wrong mappings by finding, for each label, which is the class which is most often the class with second highest votes.
        
        \fbox{
    \begin{minipage}{0.8\textwidth}
        Please classify the given text from news articles by topic as  3 (Sports) or 0 (World) or 3 (Business) or 2 (Science or Technology). Don't try to answer any question or respond to the text. Your output should only be a number from {0,1,2,3}.
    
    Text: query
    
    Answer: mapped label
    
    Text: query(i) 
    
    Answer:
    \end{minipage}
    }
        
    \end{itemize}
\end{enumerate}

\subsection{Private-kNN}

For Private-kNN, we work with the CIFAR10, Fashion MNIST, SST2 and AGNEWS datasets. We use the ViT-L/16 \citep{dosovitskiy2020image} models pretrained on ImageNet21k \citep{deng2009imagenet} to extract the features for image classification datasets and the RoBERTa-Large \citep{liu2019roberta} model to extract the features for the text classification datasets. Using these features, we train a private kNN classifier with $k=200$ and subsampling rate $\gamma=0.2$ and using Gaussian noise with standrard deviation $\sigma = 30$ for image datasets and $\sigma = 20$ for text datasets. Because the privacy analysis of Private kNN involves subsampling, we only use the data independent privacy analysis of the subsampled Gaussian mechanism as a baseline. We evaluate the privacy of Private-kNN by auditing the responses on $\numqueries = 1000$ queries.
As with other algorithms, the $\natmiadvquer$ ($\poismiadvquer$) adds a (mislabelled) test point and repeatedly queries the same point. For $\poisminatquer$, we consider three different adversaries which find a point maximizing the probability of being in the top-k for a given set of natural queries, from the test set, from the train set and from the whole feature space, respectively. To do this, we first find the value of the expected number of times a datapoint would show up as a vote contributing teacher in the top-k nearest neighbours for the whole sequence of queries. Let this expected value for point $x$ be $E(x)$. Then,
\begin{equation*}
    E(x) = \sum_{i = 1}^{\numqueries} \mathbb{P}(x\text{ is in top-}k\text{ for query } q_{i}),
\end{equation*}
where,
\begin{equation*}
    \mathbb{P}(x\text{ is in top-}k\text{ for query } q) = 
    \begin{cases}
        \gamma & \text{if } r_q \leq k, \\
        \sum_{i=0}^{k-1} \binom{r_q - 1}{i} \gamma^{i + 1} (1-\gamma)^{r_q - i - 1} & \text{if } r_q > k.
    \end{cases}
\end{equation*}

Two of the $\poisminatquer$ adversaries we consider find $E(x)$ maximizing train and test point respectively. For the embedding adversary, we come up with a heuristic to define a worst case embedding which is maximizes $E(x)$. For all the test points with a particular label, we collect the list of top-$s$ indices in a histogram. Using this histogram as weights, we combine all the embeddings in the support of this histogram to give a point. This point performs extremely well especially for text classification dataset as it gives a $E(x)$ score of almost $\gamma\numqueries$ which is its maximum attainable value. 

\section{Additional Plots}
\label{appen:add_plots}

In this section, we plot additional plots for the interested reader. For each algorithm and adversary, we plot a bar plot of the privacy leakage converted to $\adiffp$-DP against theoretical values and the performance of both the $k$-cut and $2$-cut audits. Along with this, we also plot a the Renyi DP audit plot for one dataset each as a representative for comparison.

\subsection{PATE}

\begin{figure}[!htb]
    \centering
    \begin{subfigure}[t]{0.45\textwidth}
        \centering
        \includegraphics[width=\textwidth]{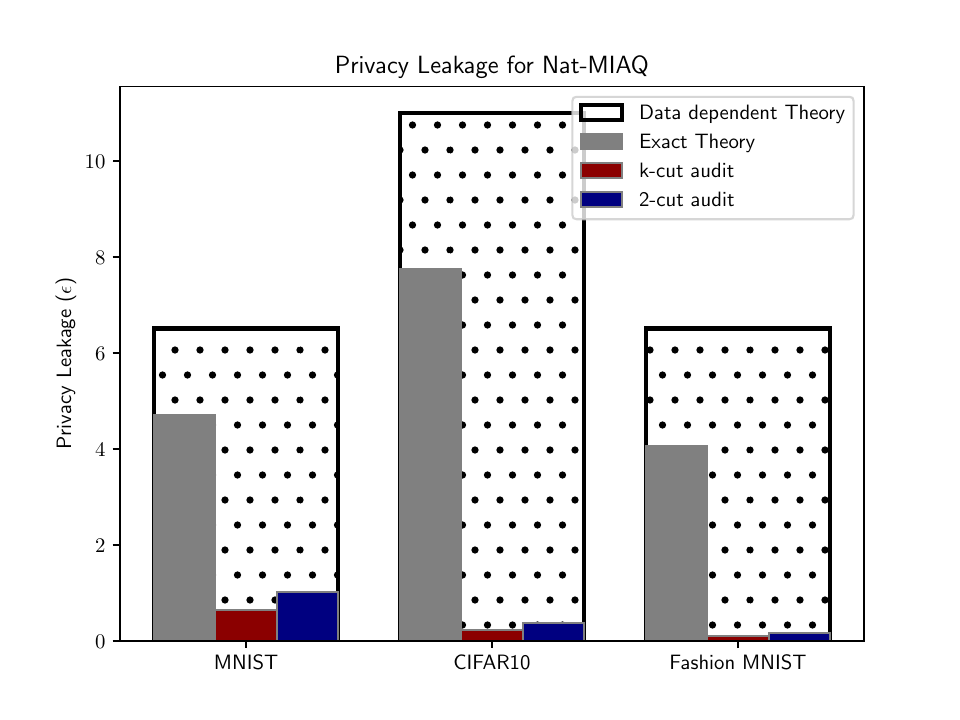}
        \caption{}
    \end{subfigure}%
    ~ 
    \begin{subfigure}[t]{0.45\textwidth}
        \centering
        \includegraphics[width=\textwidth]{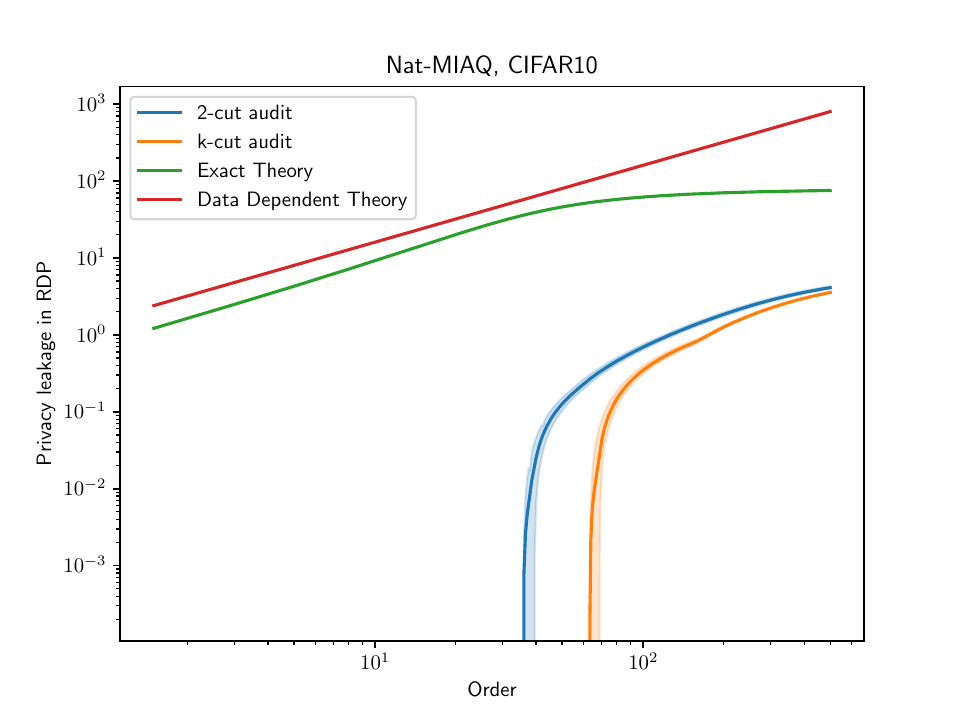}
        \caption{}
    \end{subfigure}
    \caption{Worst case for $\natmiadvquer$ adversary}
\end{figure}

\begin{figure}[!htb]
    \centering
    \begin{subfigure}[t]{0.45\textwidth}
        \centering
        \includegraphics[width=\textwidth]{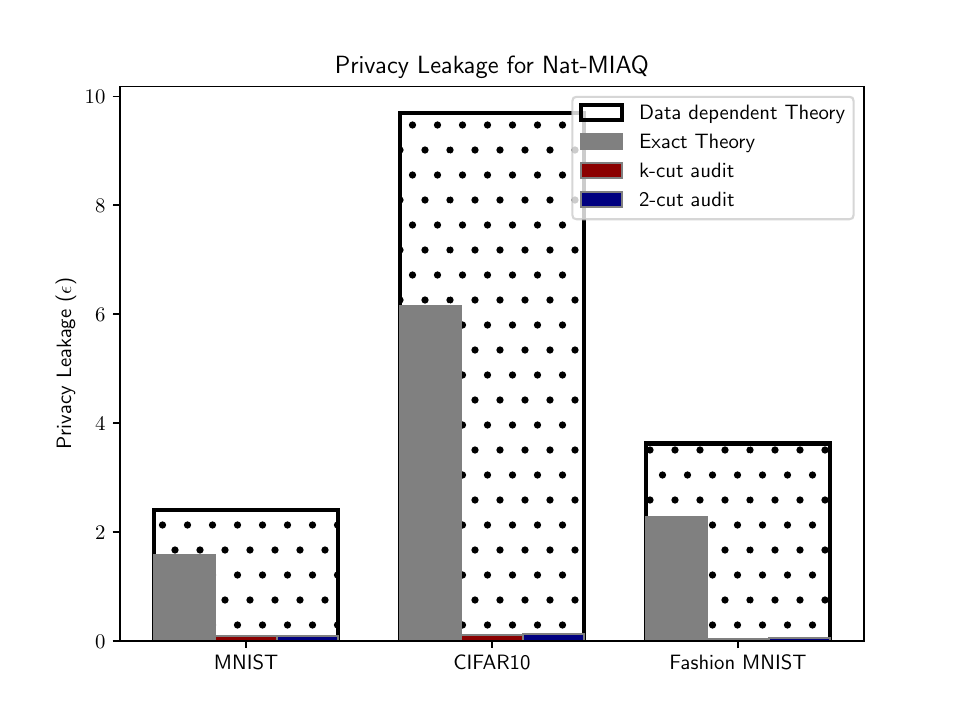}
        \caption{}
    \end{subfigure}%
    ~ 
    \begin{subfigure}[t]{0.45\textwidth}
        \centering
        \includegraphics[width=\textwidth]{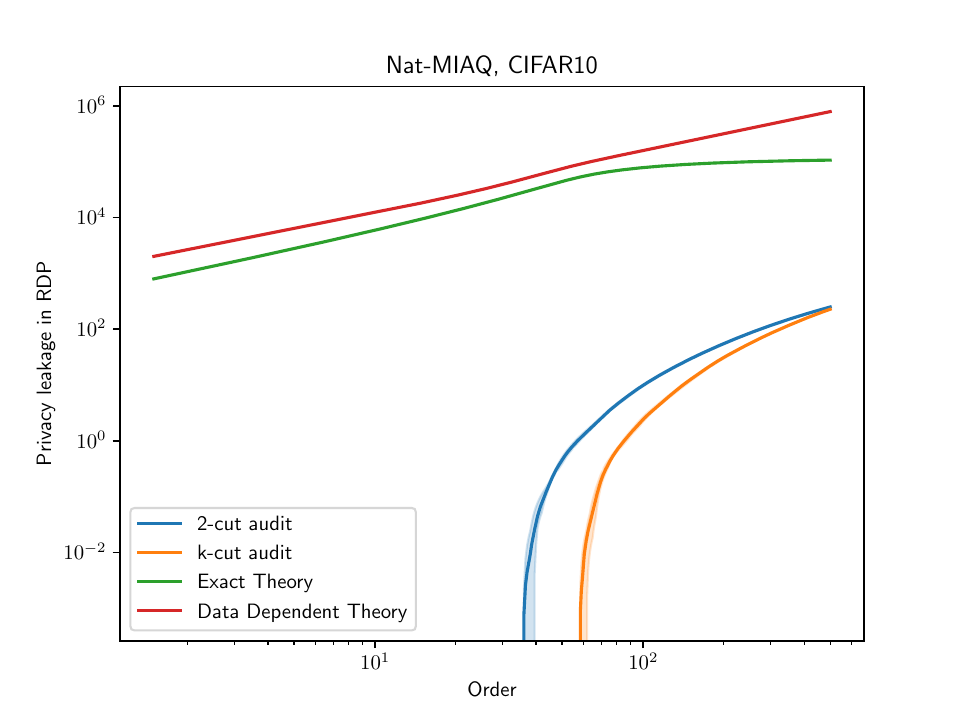}
        \caption{}
    \end{subfigure}
    \caption{Average case for $\natmiadvquer$ adversary}
\end{figure}

\begin{figure}[!htb]
    \centering
    \begin{subfigure}[t]{0.45\textwidth}
        \centering
        \includegraphics[width=\textwidth]{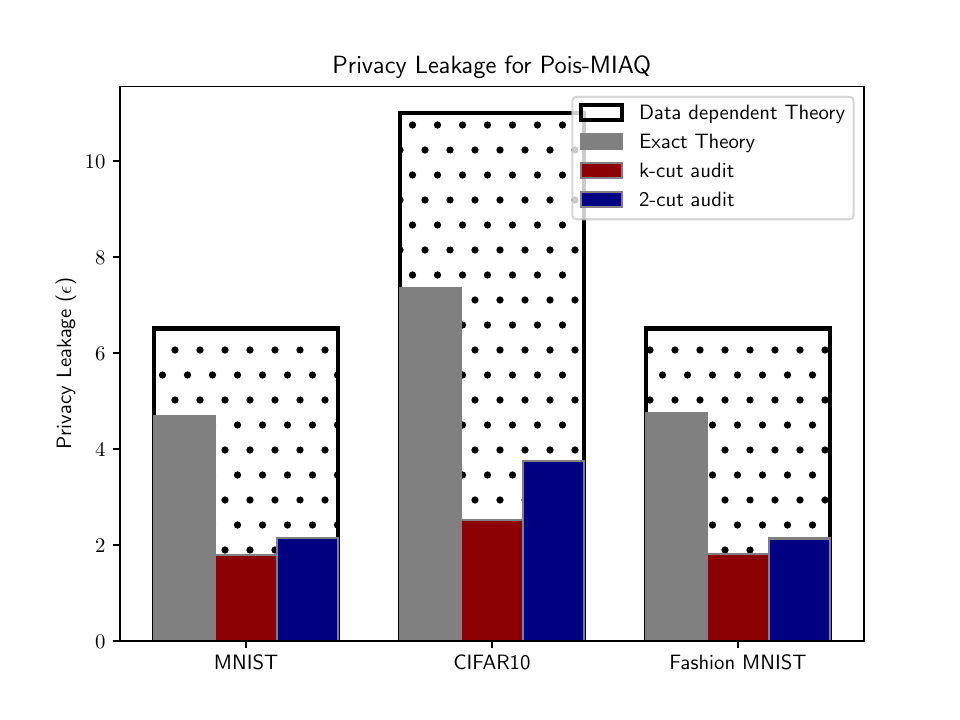}
        \caption{}
    \end{subfigure}%
    ~ 
    \begin{subfigure}[t]{0.45\textwidth}
        \centering
        \includegraphics[width=\textwidth]{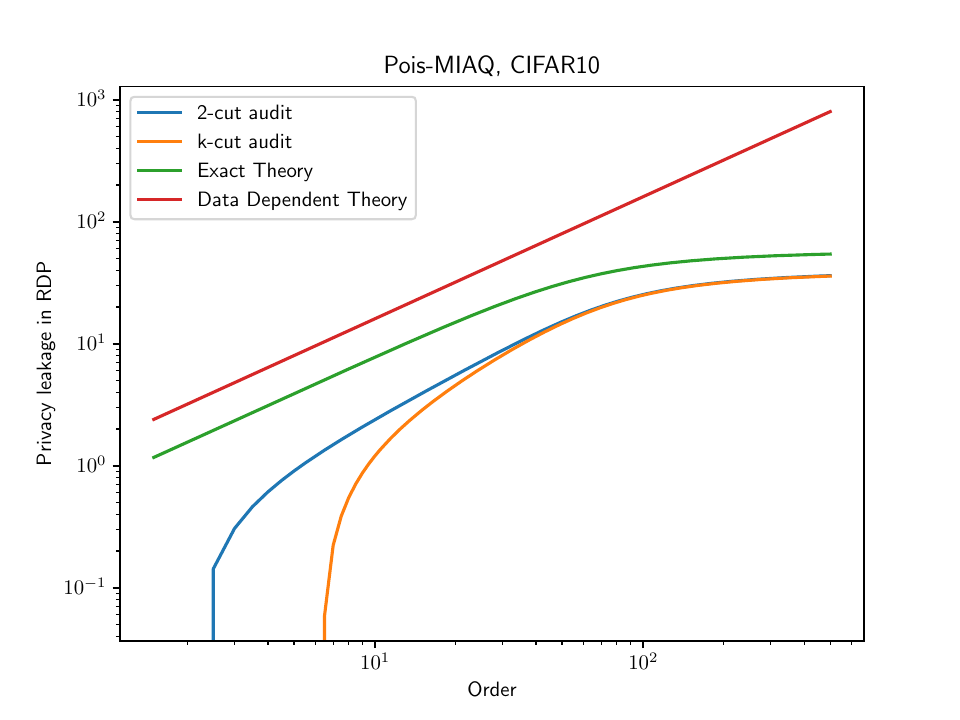}
        \caption{}
    \end{subfigure}
    \caption{Worst case for $\poismiadvquer$ adversary}
\end{figure}

\begin{figure}[!htb]
    \centering
    \begin{subfigure}[t]{0.45\textwidth}
        \centering
        \includegraphics[width=\textwidth]{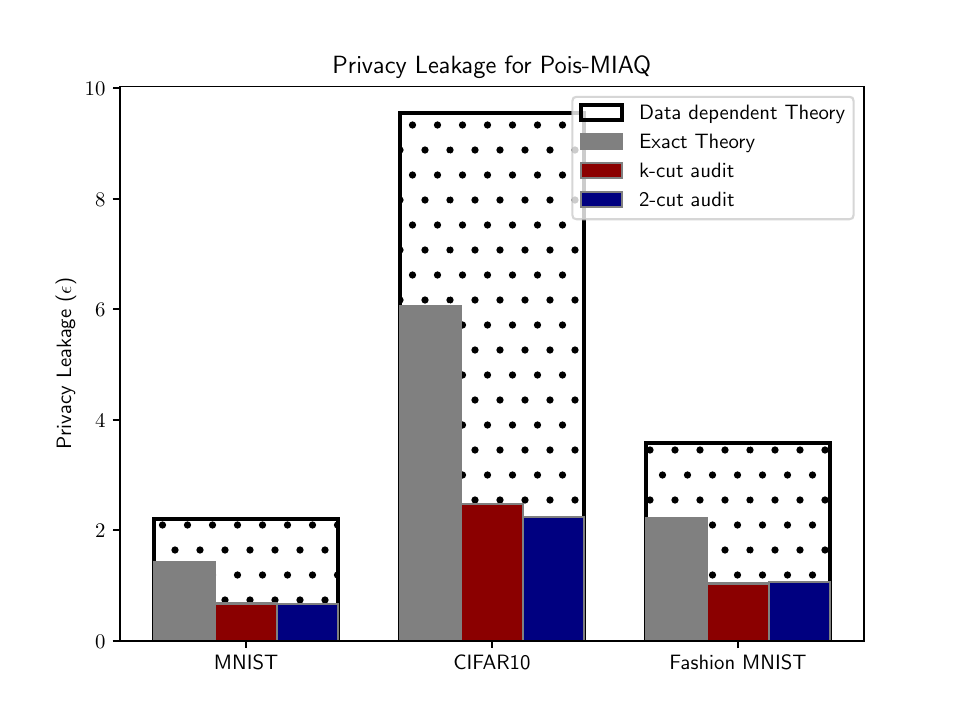}
        \caption{}
    \end{subfigure}%
    ~ 
    \begin{subfigure}[t]{0.45\textwidth}
        \centering
        \includegraphics[width=\textwidth]{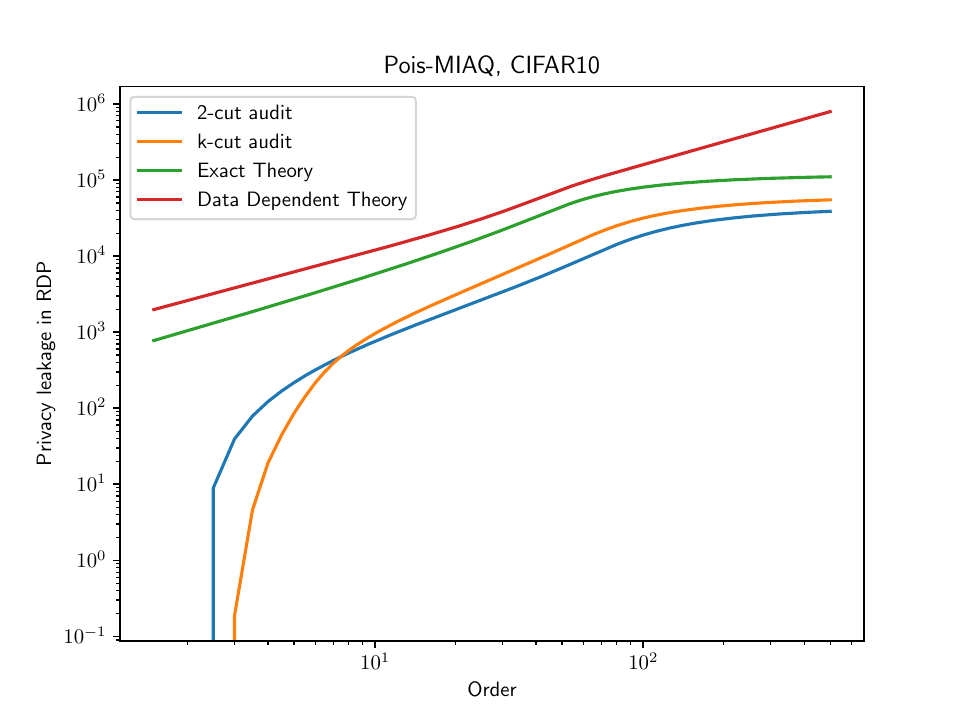}
        \caption{}
    \end{subfigure}
    \caption{Average case for $\poismiadvquer$ adversary}
\end{figure}

\begin{figure}[!htb]
    \centering
    \begin{subfigure}[t]{0.45\textwidth}
        \centering
        \includegraphics[width=\textwidth]{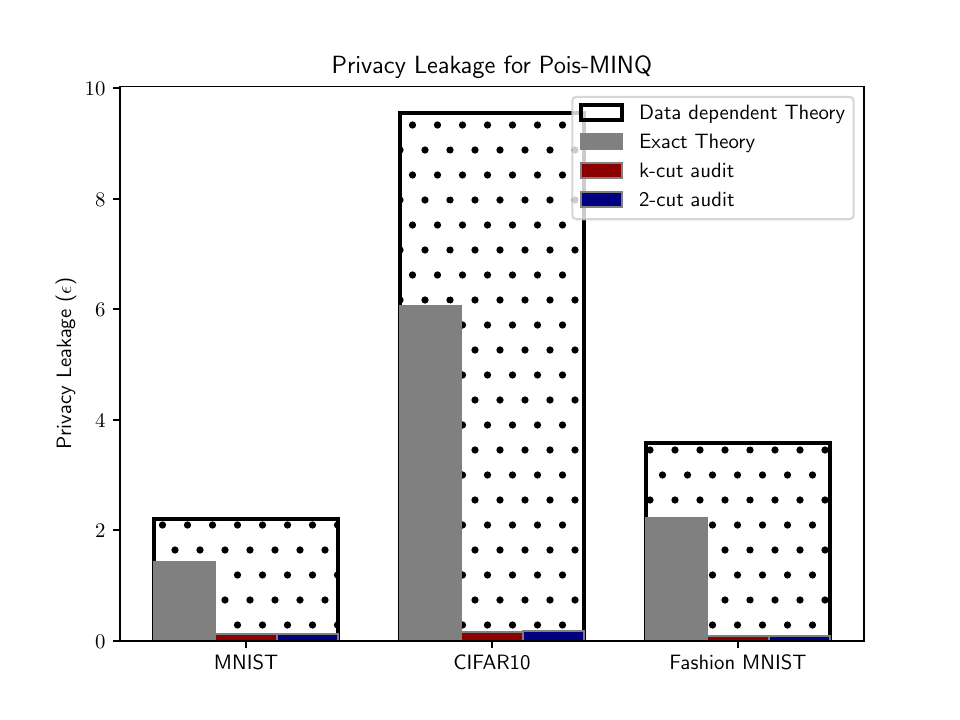}
        \caption{}
    \end{subfigure}%
    ~ 
    \begin{subfigure}[t]{0.45\textwidth}
        \centering
        \includegraphics[width=\textwidth]{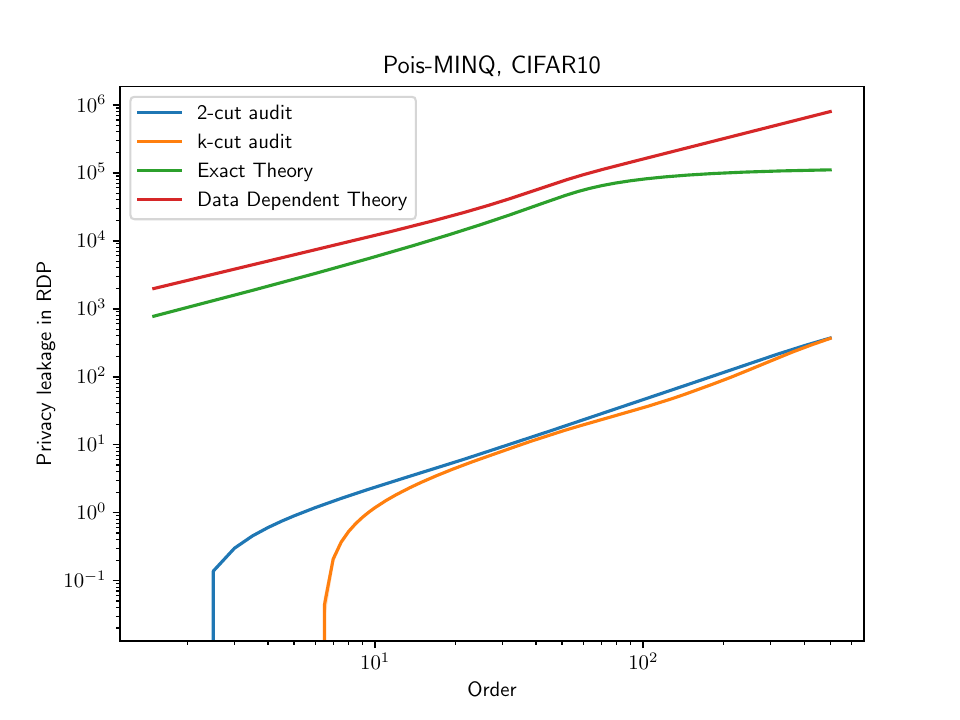}
        \caption{}
    \end{subfigure}
    \caption{$\poisminatquer$ adversary}
\end{figure}

\newpage

\subsection{CaPC}

\begin{figure}[!htb]
    \centering
    \begin{subfigure}[t]{0.45\textwidth}
        \centering
        \includegraphics[width=\textwidth]{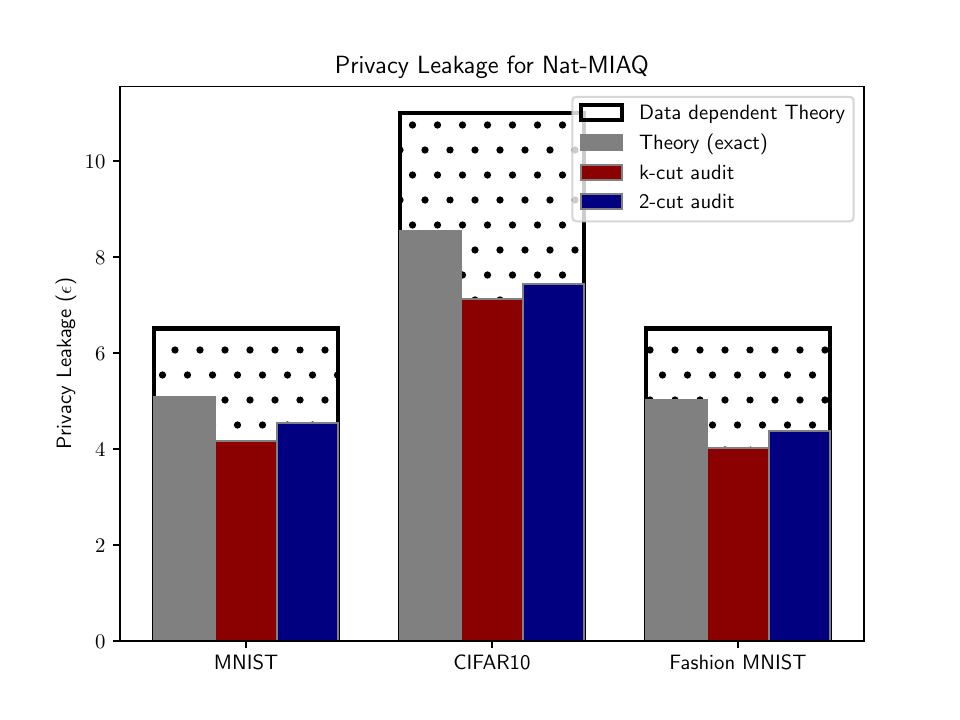}
        \caption{}
    \end{subfigure}%
    ~ 
    \begin{subfigure}[t]{0.45\textwidth}
        \centering
        \includegraphics[width=\textwidth]{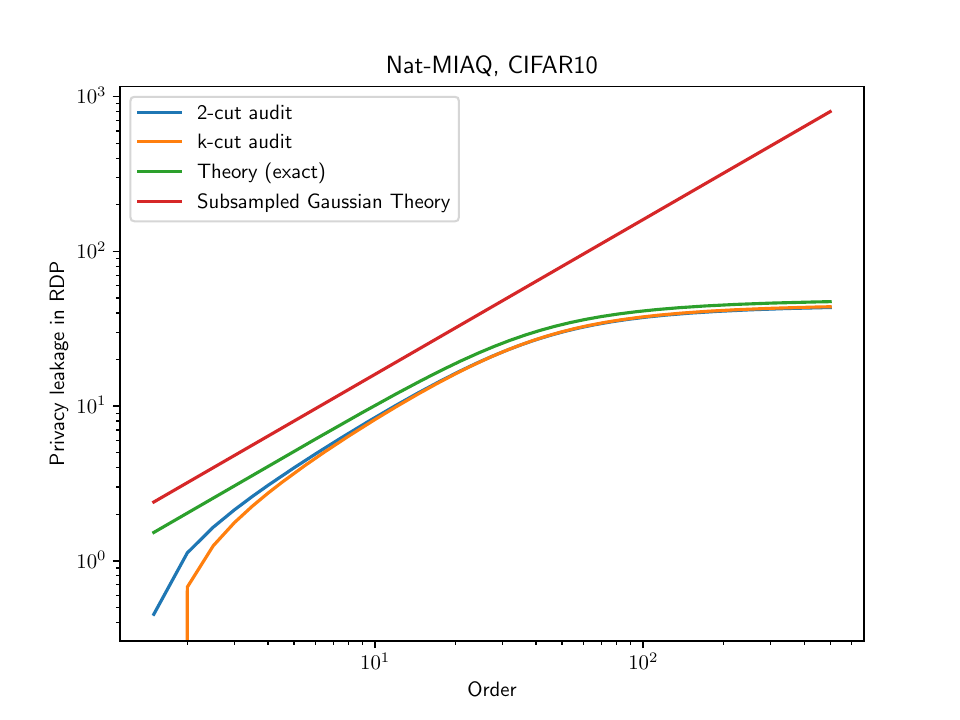}
        \caption{}
    \end{subfigure}
    \caption{Worst case for $\natmiadvquer$ adversary}
\end{figure}

\begin{figure}[!htb]
    \centering
    \begin{subfigure}[t]{0.45\textwidth}
        \centering
        \includegraphics[width=\textwidth]{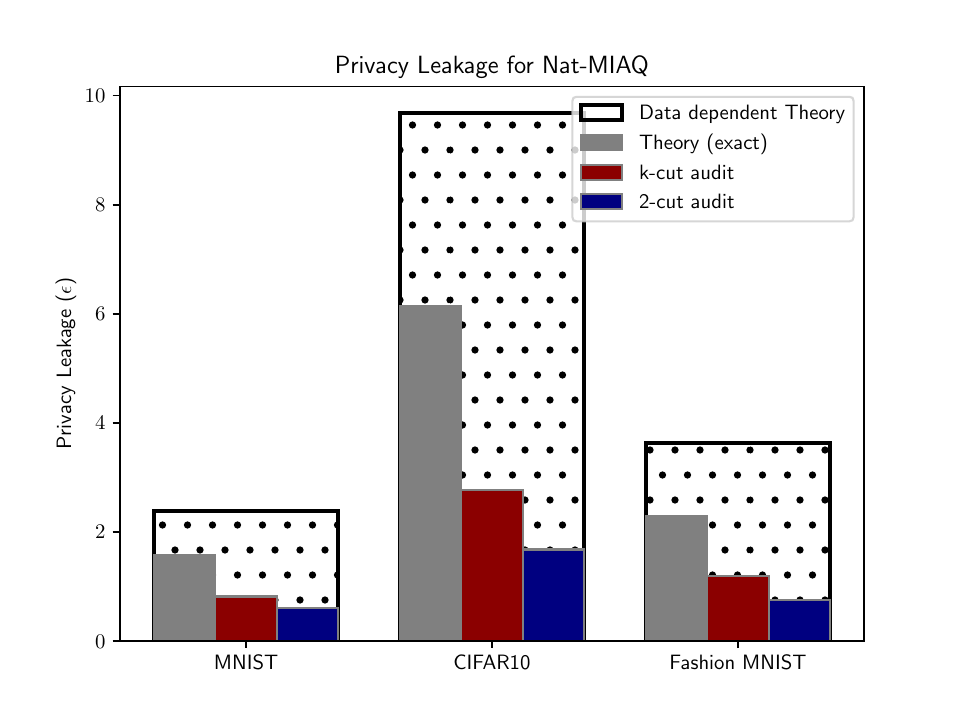}
        \caption{}
    \end{subfigure}%
    ~ 
    \begin{subfigure}[t]{0.45\textwidth}
        \centering
        \includegraphics[width=\textwidth]{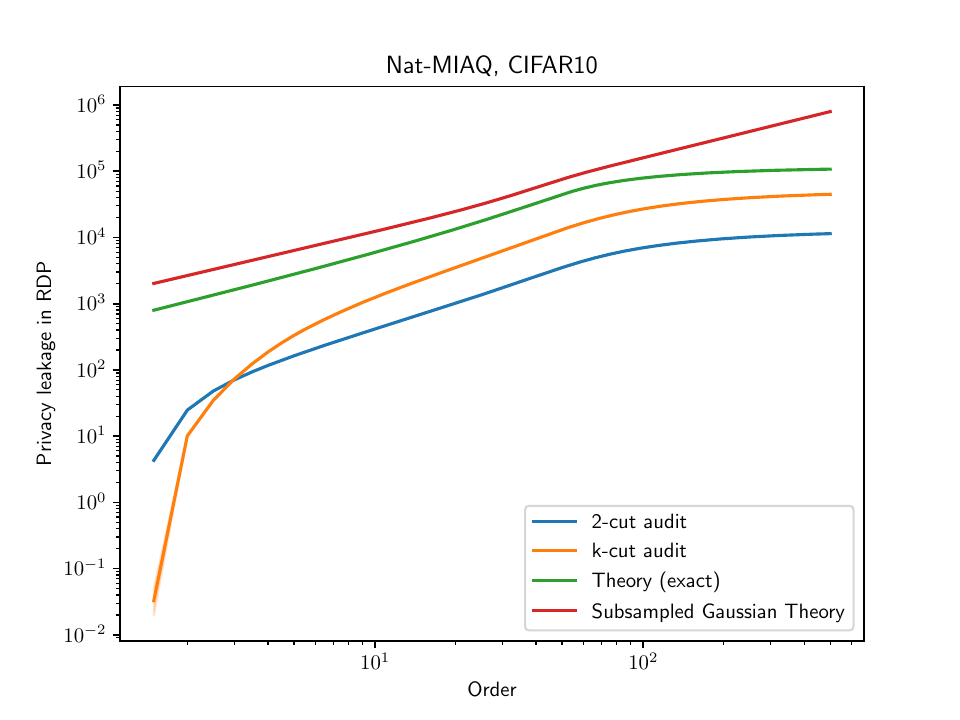}
        \caption{}
    \end{subfigure}
    \caption{Average case for $\natmiadvquer$ adversary}
\end{figure}

\begin{figure}[!htb]
    \centering
    \begin{subfigure}[t]{0.45\textwidth}
        \centering
        \includegraphics[width=\textwidth]{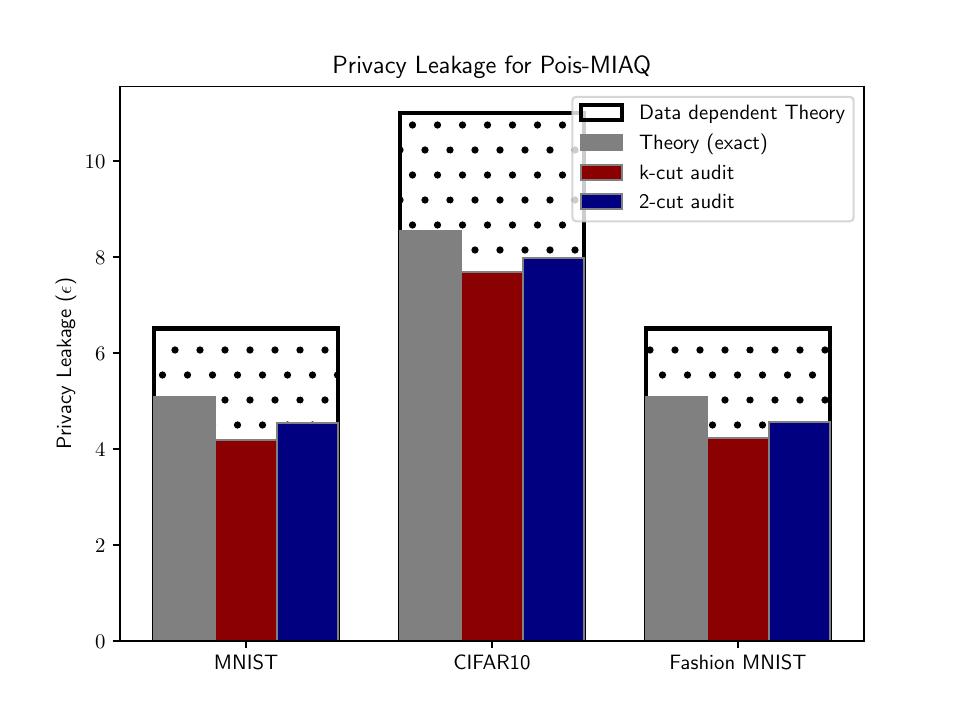}
        \caption{}
    \end{subfigure}%
    ~ 
    \begin{subfigure}[t]{0.45\textwidth}
        \centering
        \includegraphics[width=\textwidth]{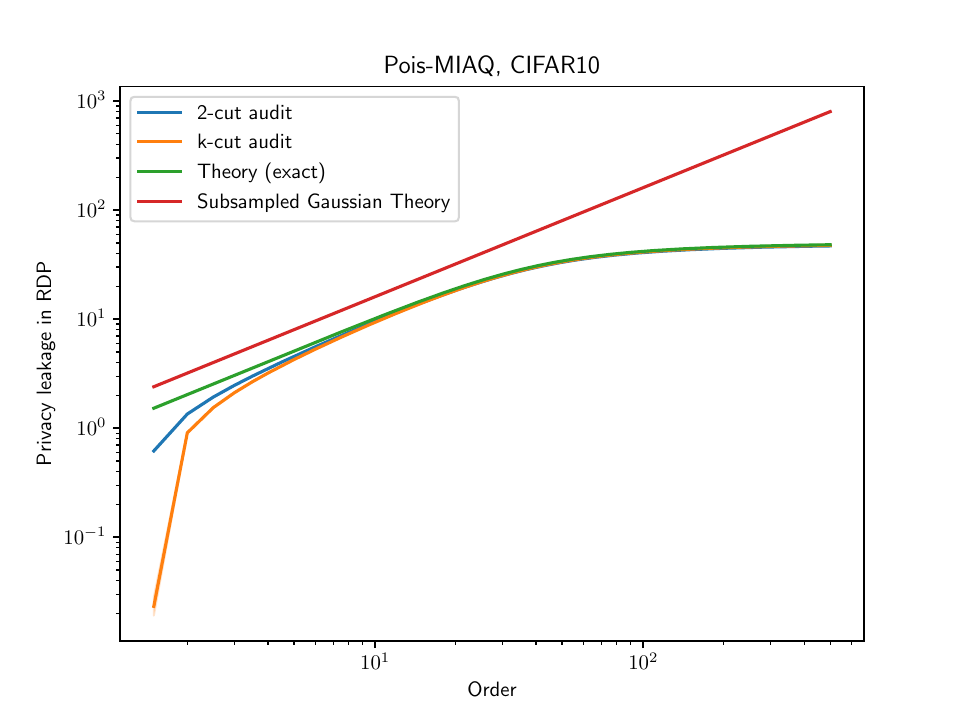}
        \caption{}
    \end{subfigure}
    \caption{Worst case for $\poismiadvquer$ adversary}
\end{figure}

\begin{figure}[!htb]
    \centering
    \begin{subfigure}[t]{0.45\textwidth}
        \centering
        \includegraphics[width=\textwidth]{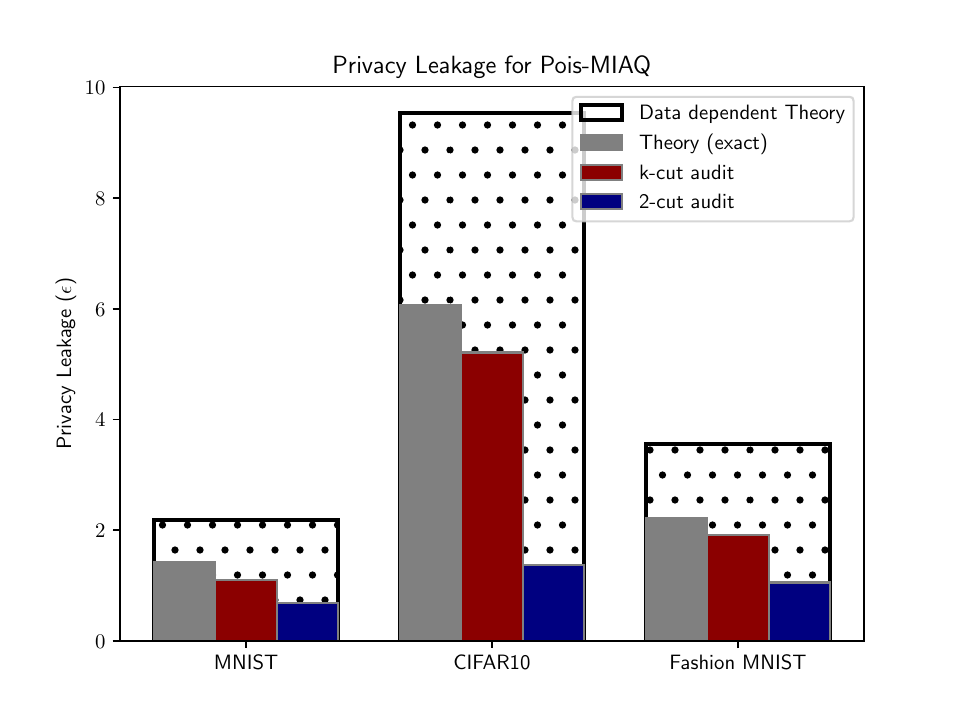}
        \caption{}
    \end{subfigure}%
    ~ 
    \begin{subfigure}[t]{0.45\textwidth}
        \centering
        \includegraphics[width=\textwidth]{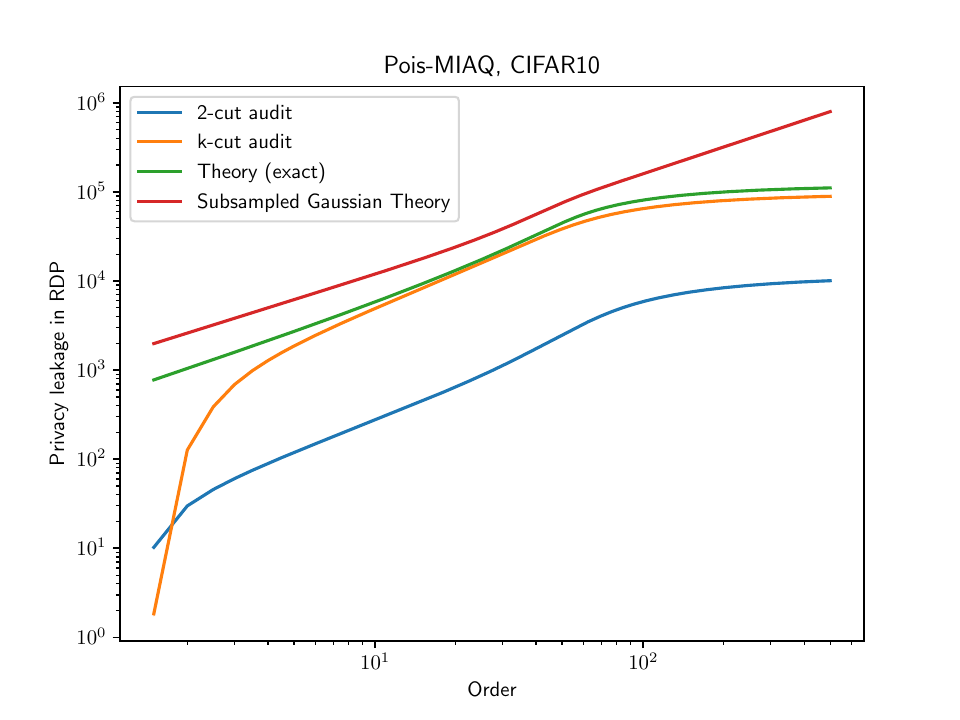}
        \caption{}
    \end{subfigure}
    \caption{Average case for $\poismiadvquer$ adversary}
\end{figure}

\begin{figure}[!htb]
    \centering
    \begin{subfigure}[t]{0.45\textwidth}
        \centering
        \includegraphics[width=\textwidth]{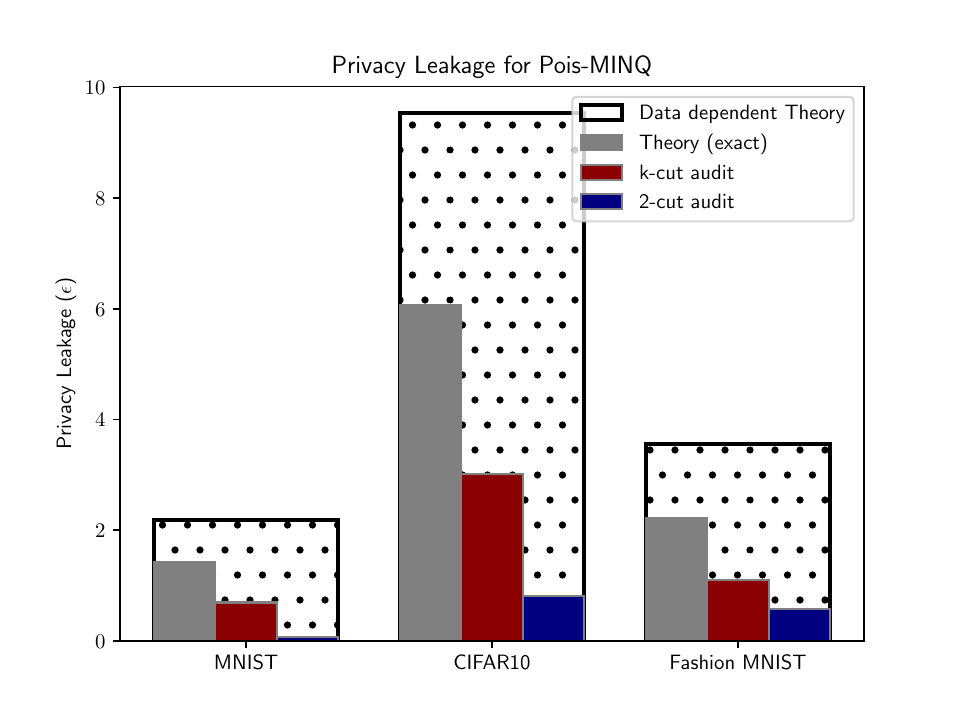}
        \caption{}
    \end{subfigure}%
    ~ 
    \begin{subfigure}[t]{0.45\textwidth}
        \centering
        \includegraphics[width=\textwidth]{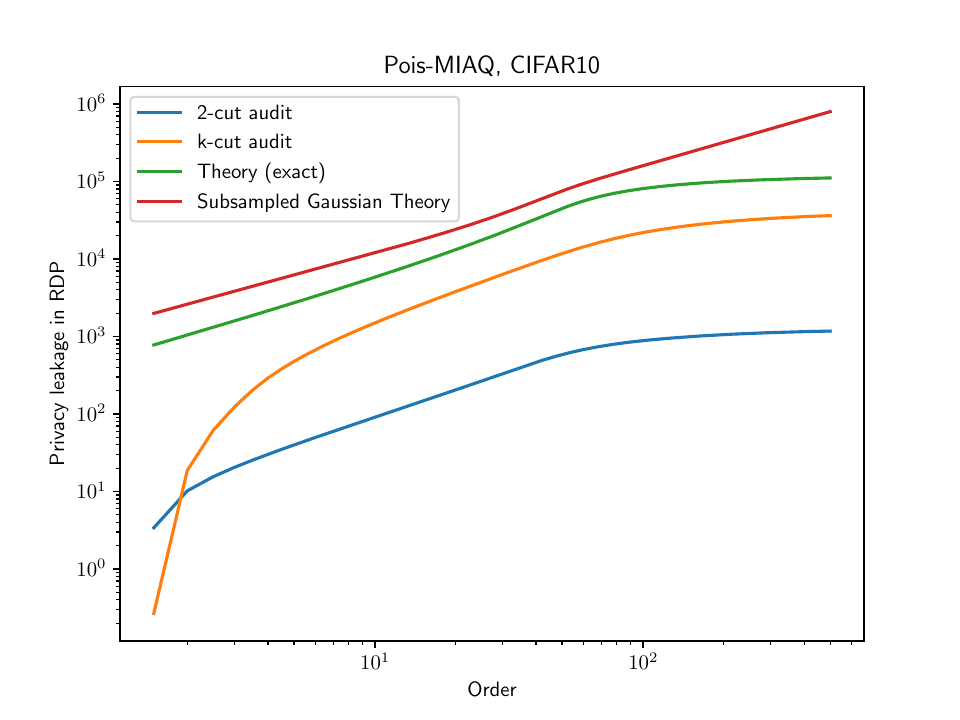}
        \caption{}
    \end{subfigure}
    \caption{$\poisminatquer$ adversary}
\end{figure}

\newpage
\subsection{PromptPATE}

\begin{figure}[!htb]
    \centering
    \begin{subfigure}[t]{0.45\textwidth}
        \centering
        \includegraphics[width=\textwidth]{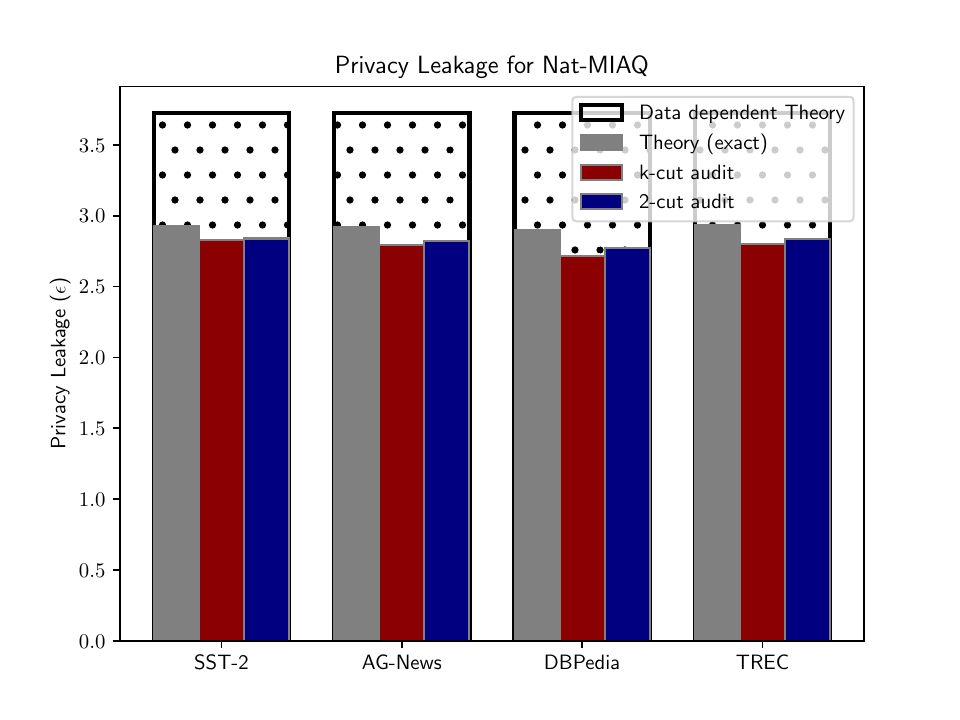}
        \caption{}
    \end{subfigure}%
    ~ 
    \begin{subfigure}[t]{0.45\textwidth}
        \centering
        \includegraphics[width=\textwidth]{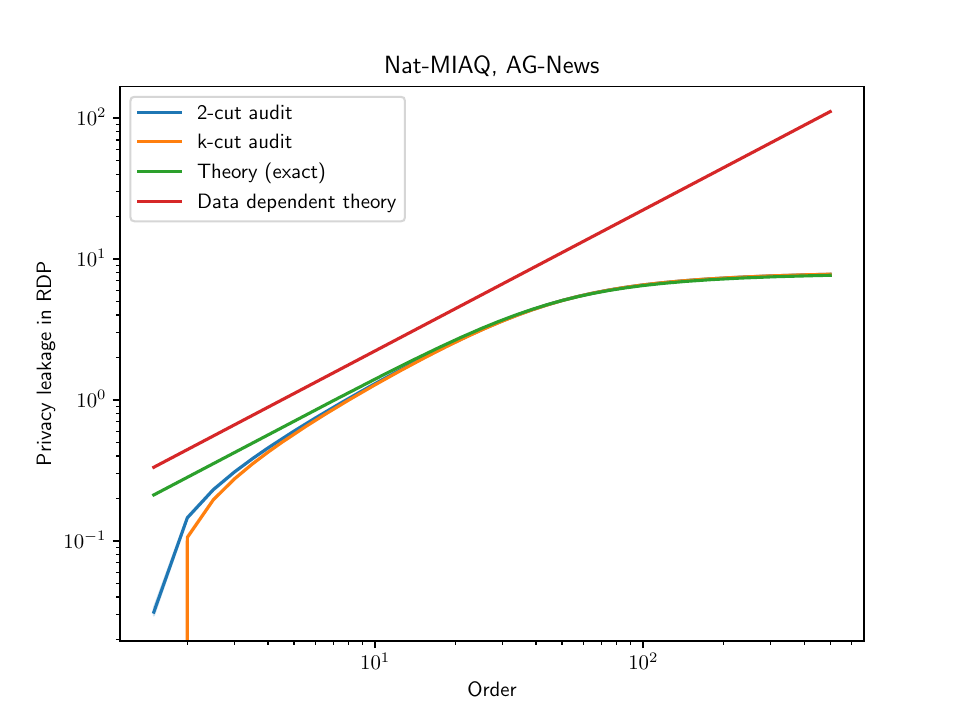}
        \caption{}
    \end{subfigure}
    \caption{Worst case for $\natmiadvquer$ adversary}
\end{figure}

\begin{figure}[!htb]
    \centering
    \begin{subfigure}[t]{0.45\textwidth}
        \centering
        \includegraphics[width=\textwidth]{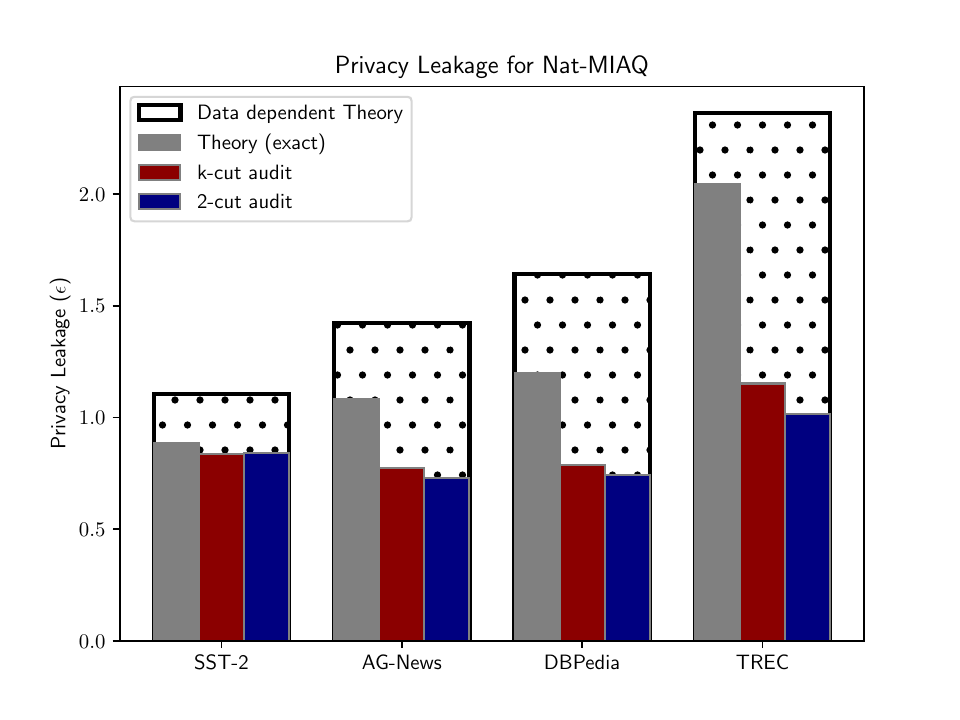}
        \caption{}
    \end{subfigure}%
    ~ 
    \begin{subfigure}[t]{0.45\textwidth}
        \centering
        \includegraphics[width=\textwidth]{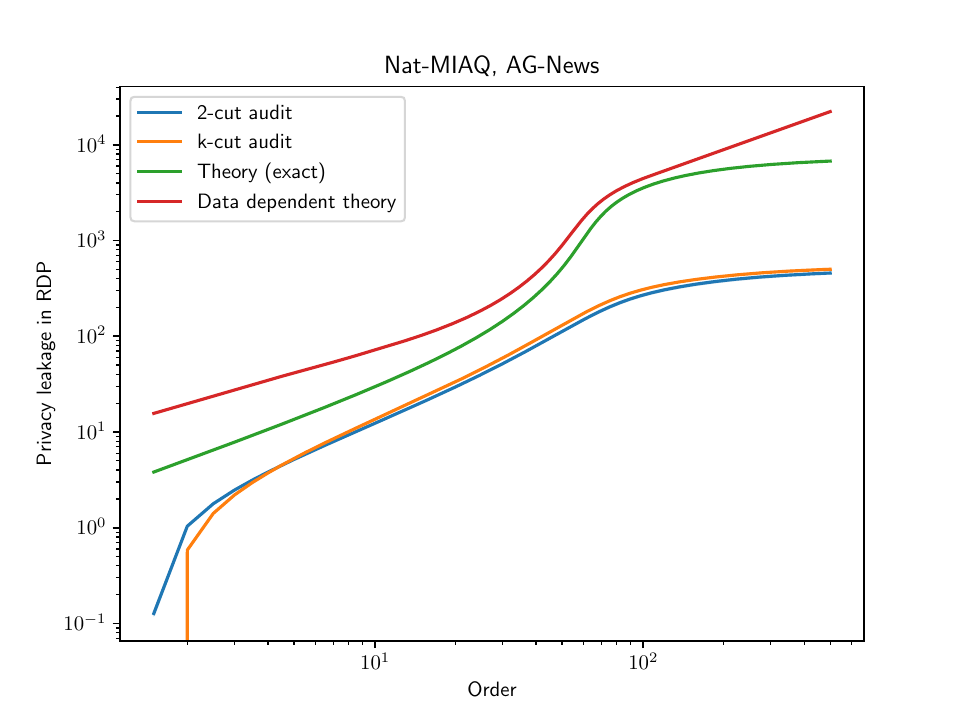}
        \caption{}
    \end{subfigure}
    \caption{Average case for $\natmiadvquer$ adversary}
\end{figure}

\begin{figure}[!htb]
    \centering
    \begin{subfigure}[t]{0.45\textwidth}
        \centering
        \includegraphics[width=\textwidth]{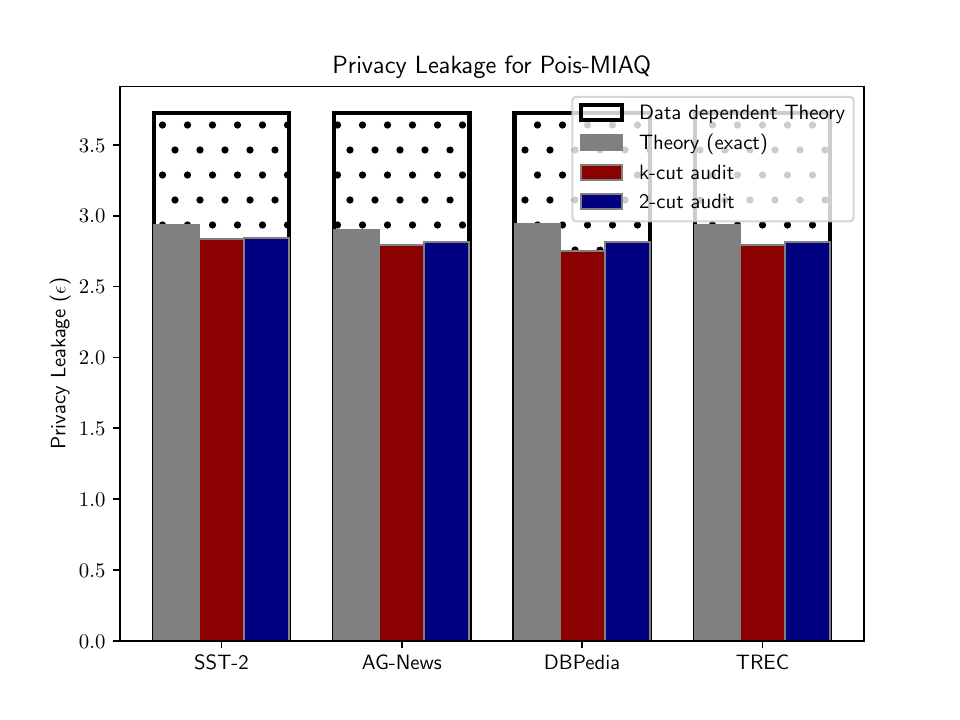}
        \caption{}
    \end{subfigure}%
    ~ 
    \begin{subfigure}[t]{0.45\textwidth}
        \centering
        \includegraphics[width=\textwidth]{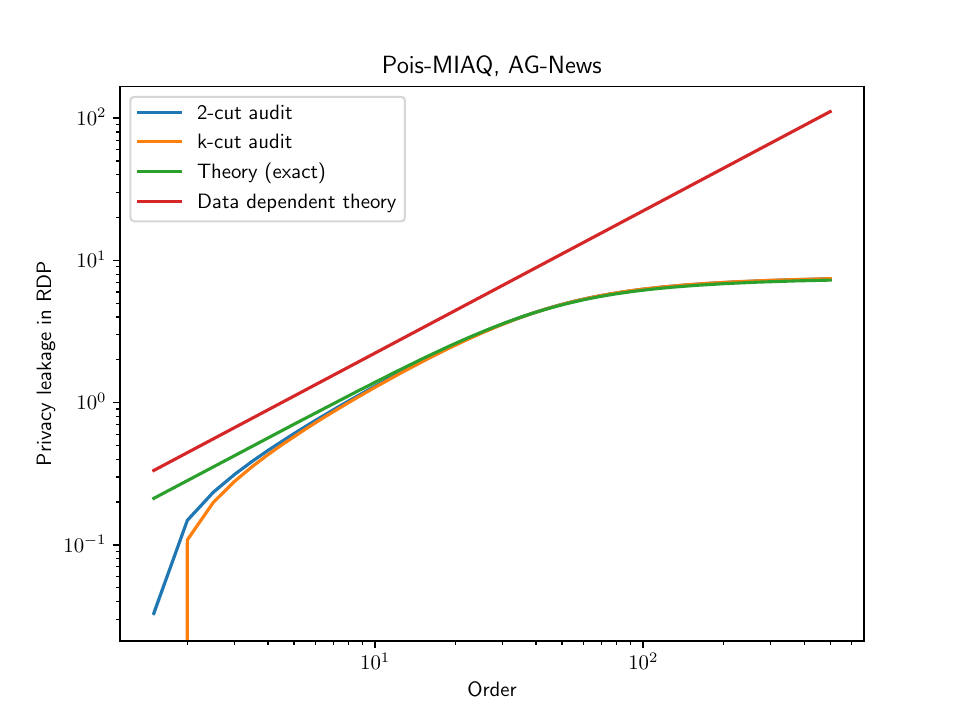}
        \caption{}
    \end{subfigure}
    \caption{Worst case for $\poismiadvquer$ adversary}
\end{figure}

\begin{figure}[!htb]
    \centering
    \begin{subfigure}[t]{0.45\textwidth}
        \centering
        \includegraphics[width=\textwidth]{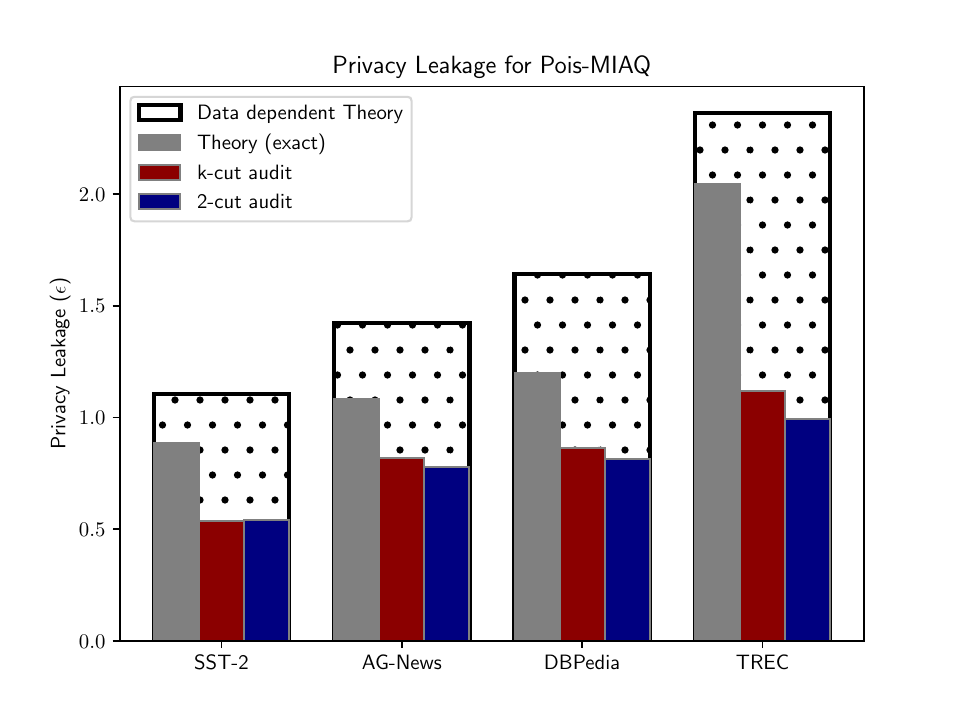}
        \caption{}
    \end{subfigure}%
    ~ 
    \begin{subfigure}[t]{0.45\textwidth}
        \centering
        \includegraphics[width=\textwidth]{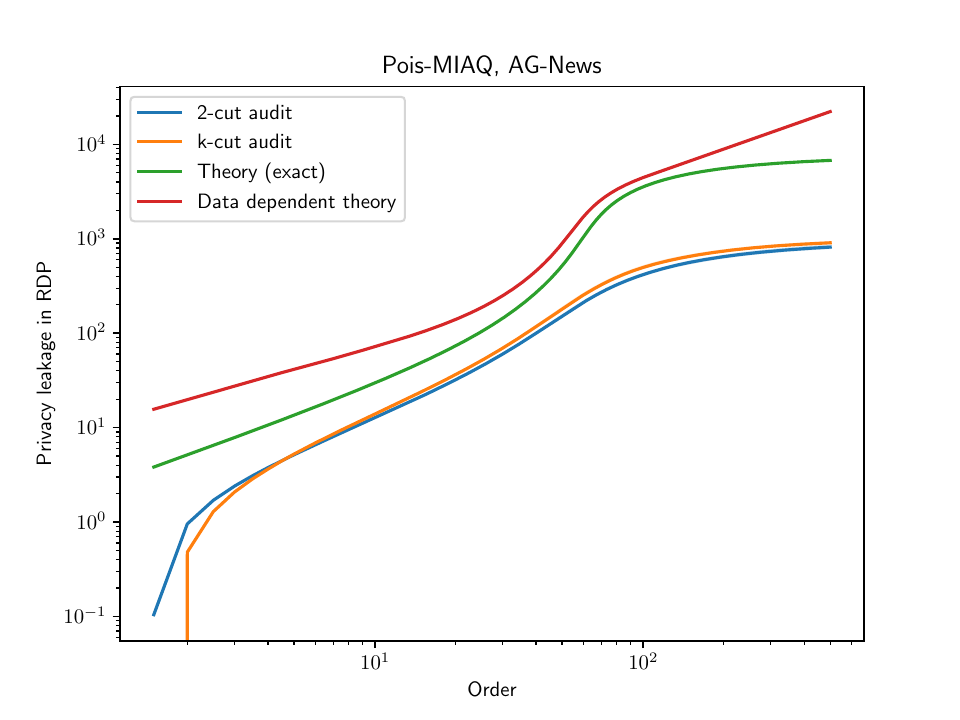}
        \caption{}
    \end{subfigure}
    \caption{Average case for $\poismiadvquer$ adversary}
\end{figure}

\begin{figure}[!htb]
    \centering
    \begin{subfigure}[t]{0.45\textwidth}
        \centering
        \includegraphics[width=\textwidth]{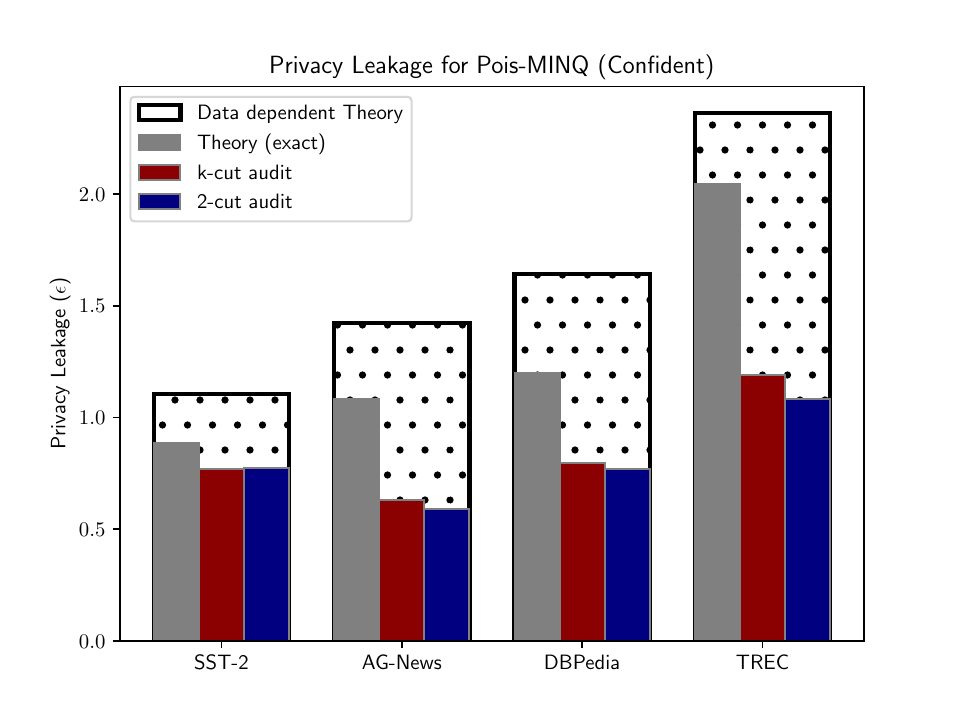}
        \caption{}
    \end{subfigure}%
    ~ 
    \begin{subfigure}[t]{0.45\textwidth}
        \centering
        \includegraphics[width=\textwidth]{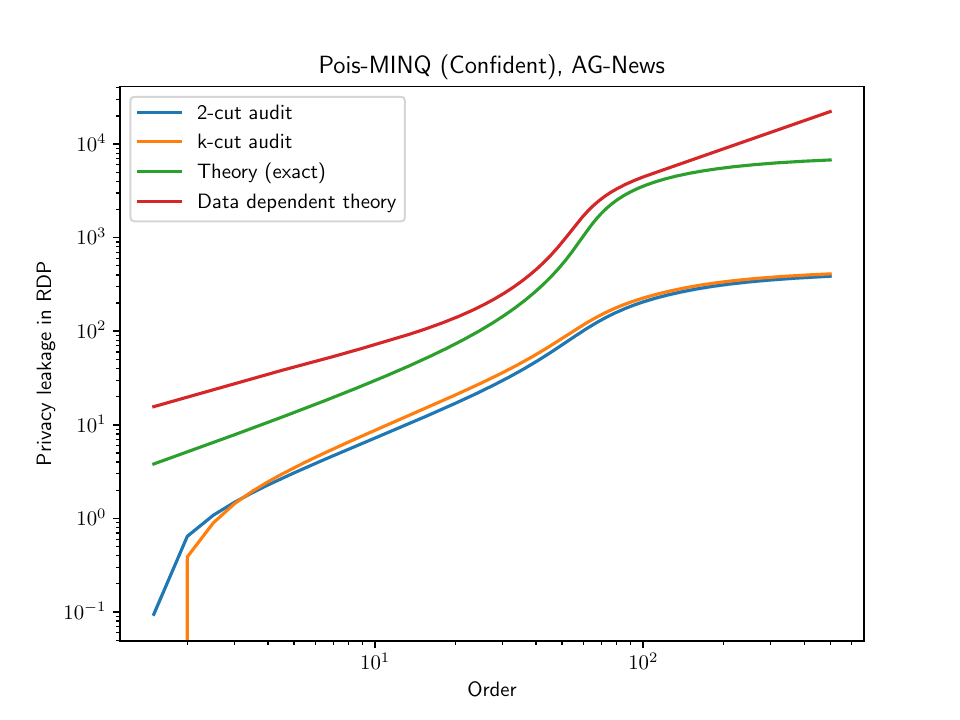}
        \caption{}
    \end{subfigure}
    \caption{$\poisminatquer$ Confident adversary}
\end{figure}

\begin{figure}[!htb]
    \centering
    \begin{subfigure}[t]{0.45\textwidth}
        \centering
        \includegraphics[width=\textwidth]{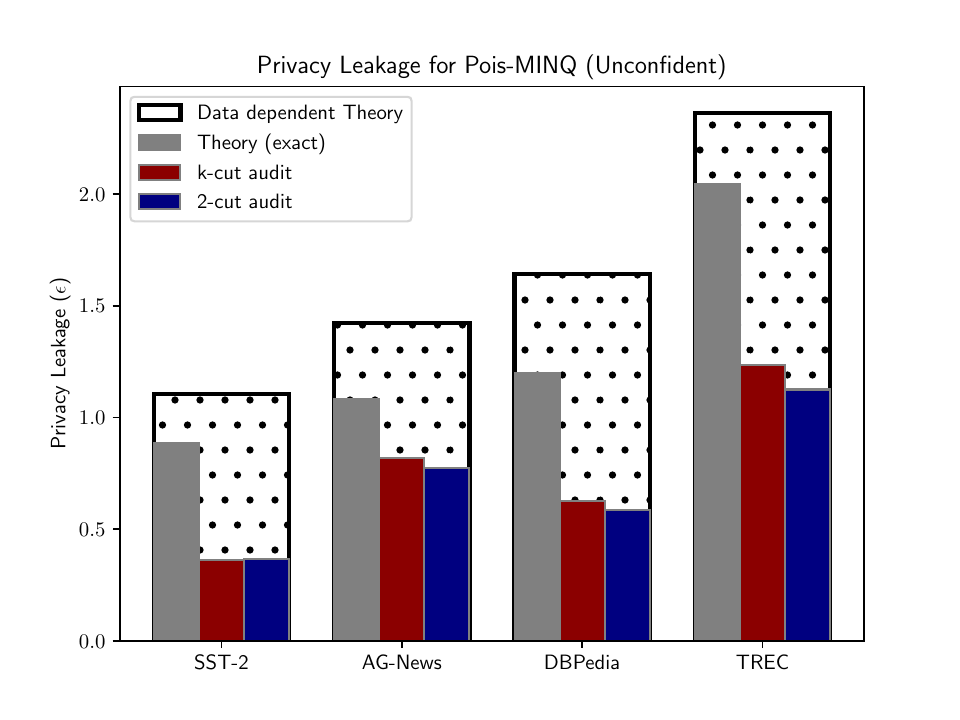}
        \caption{}
    \end{subfigure}%
    ~ 
    \begin{subfigure}[t]{0.45\textwidth}
        \centering
        \includegraphics[width=\textwidth]{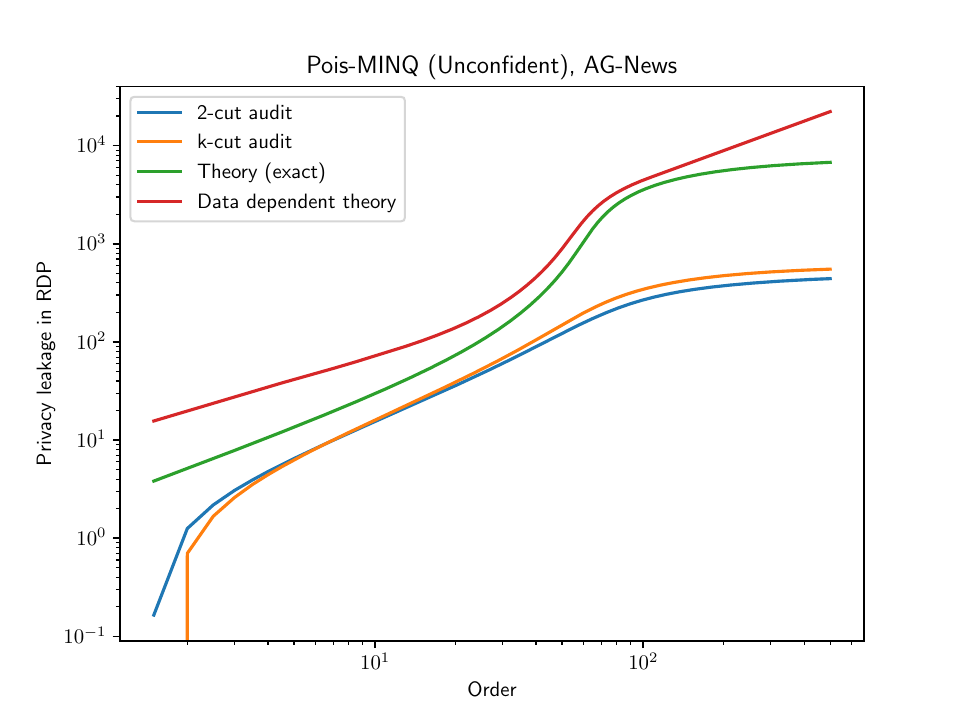}
        \caption{}
    \end{subfigure}
    \caption{$\poisminatquer$ unconfident adversary}
\end{figure}

\begin{figure}[!htb]
    \centering
    \begin{subfigure}[t]{0.45\textwidth}
        \centering
        \includegraphics[width=\textwidth]{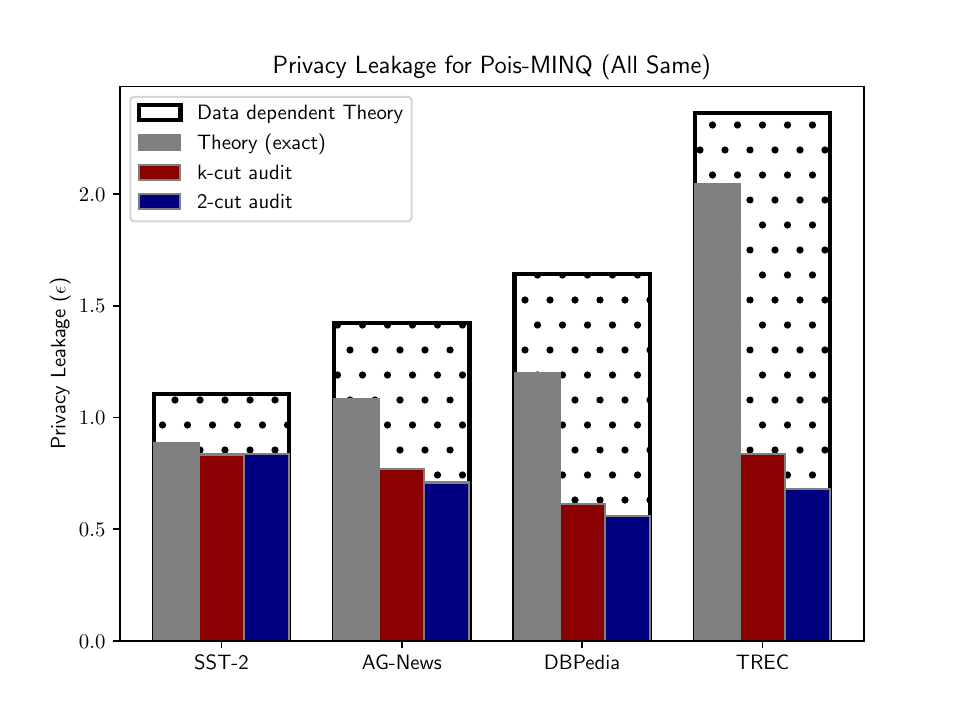}
        \caption{}
    \end{subfigure}%
    ~ 
    \begin{subfigure}[t]{0.45\textwidth}
        \centering
        \includegraphics[width=\textwidth]{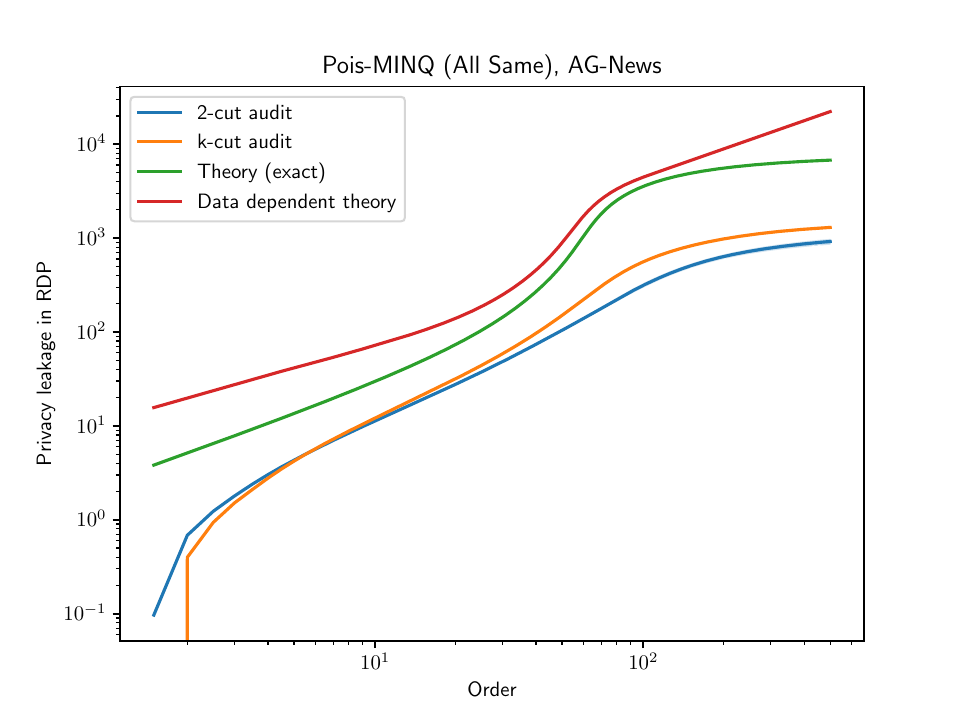}
        \caption{}
    \end{subfigure}
    \caption{$\poisminatquer$ all same adversary}
\end{figure}

\begin{figure}[!htb]
    \centering
    \begin{subfigure}[t]{0.45\textwidth}
        \centering
        \includegraphics[width=\textwidth]{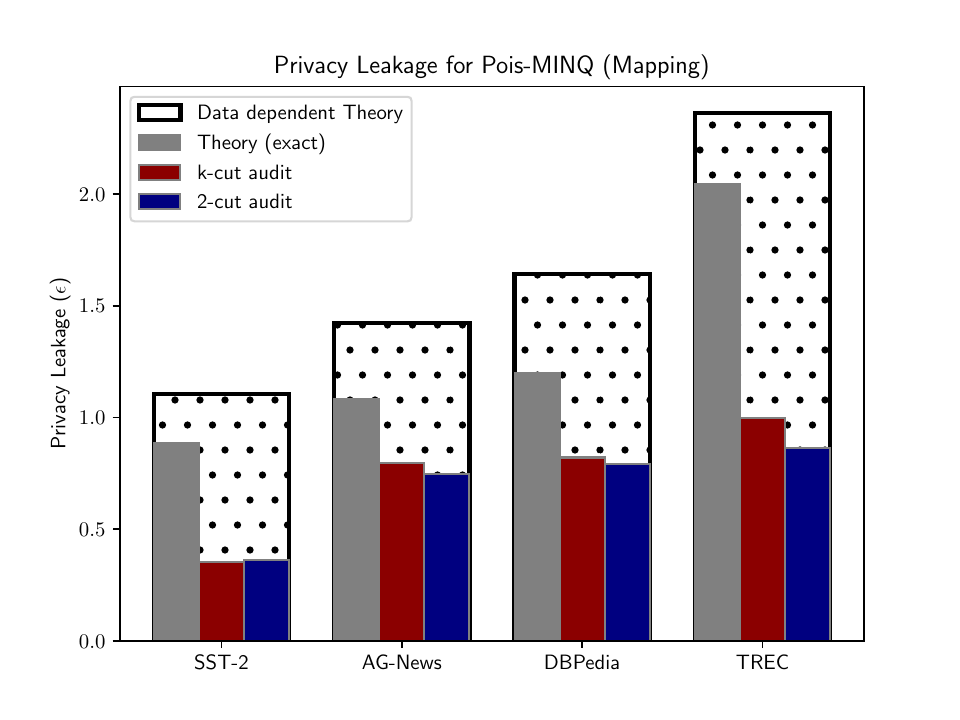}
        \caption{}
    \end{subfigure}%
    ~ 
    \begin{subfigure}[t]{0.45\textwidth}
        \centering
        \includegraphics[width=\textwidth]{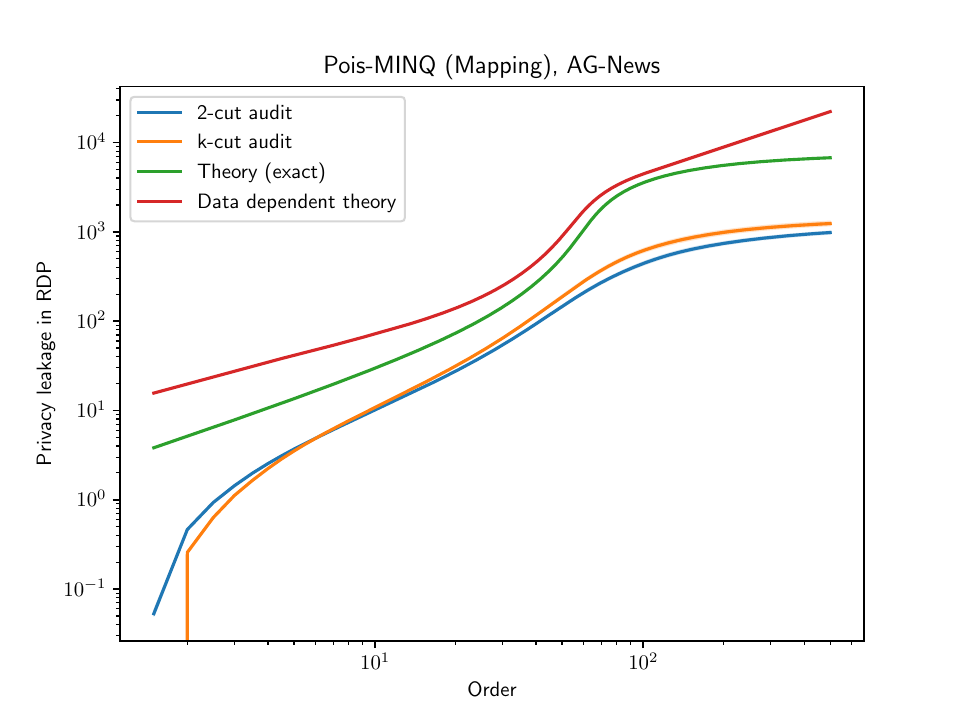}
        \caption{}
    \end{subfigure}
    \caption{$\poisminatquer$ mapping adversary}
\end{figure}

\newpage
\subsection{Private-kNN}

\begin{figure}[!htb]
    \centering
    \begin{subfigure}[t]{0.45\textwidth}
        \centering
        \includegraphics[width=\textwidth]{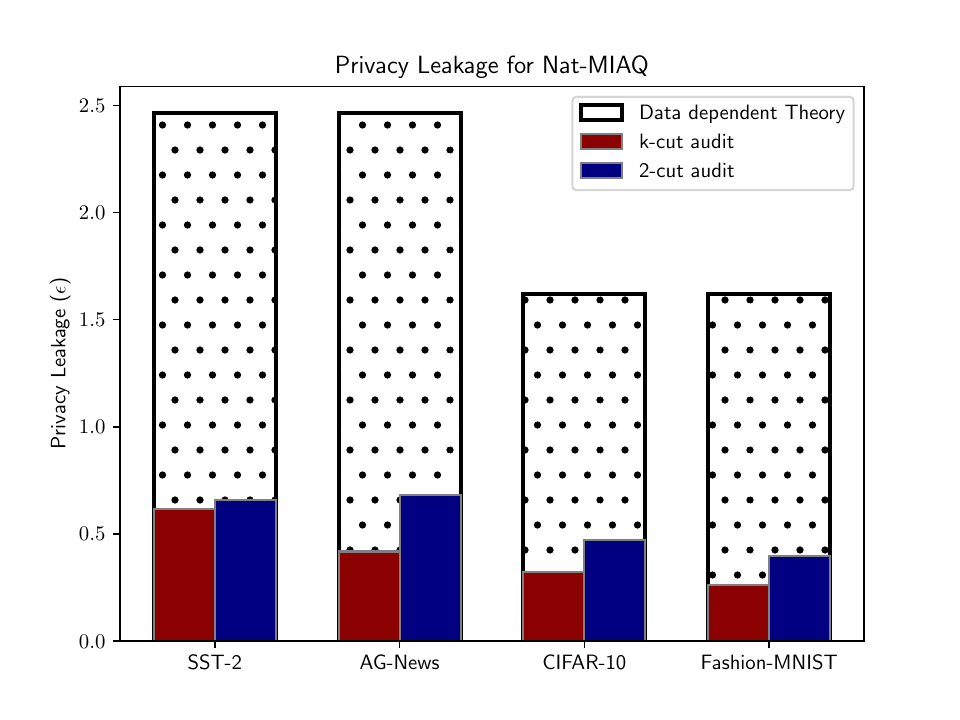}
        \caption{}
    \end{subfigure}%
    ~ 
    \begin{subfigure}[t]{0.45\textwidth}
        \centering
        \includegraphics[width=\textwidth]{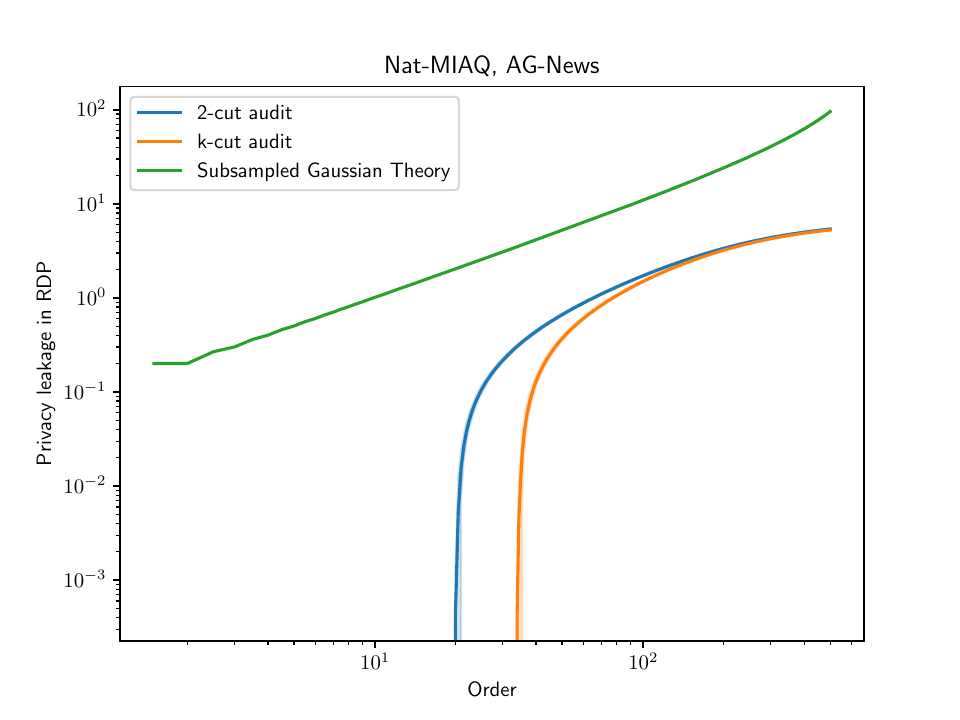}
        \caption{}
    \end{subfigure}
    \caption{Worst case for $\natmiadvquer$ adversary}
\end{figure}

\begin{figure}[!htb]
    \centering
    \begin{subfigure}[t]{0.45\textwidth}
        \centering
        \includegraphics[width=\textwidth]{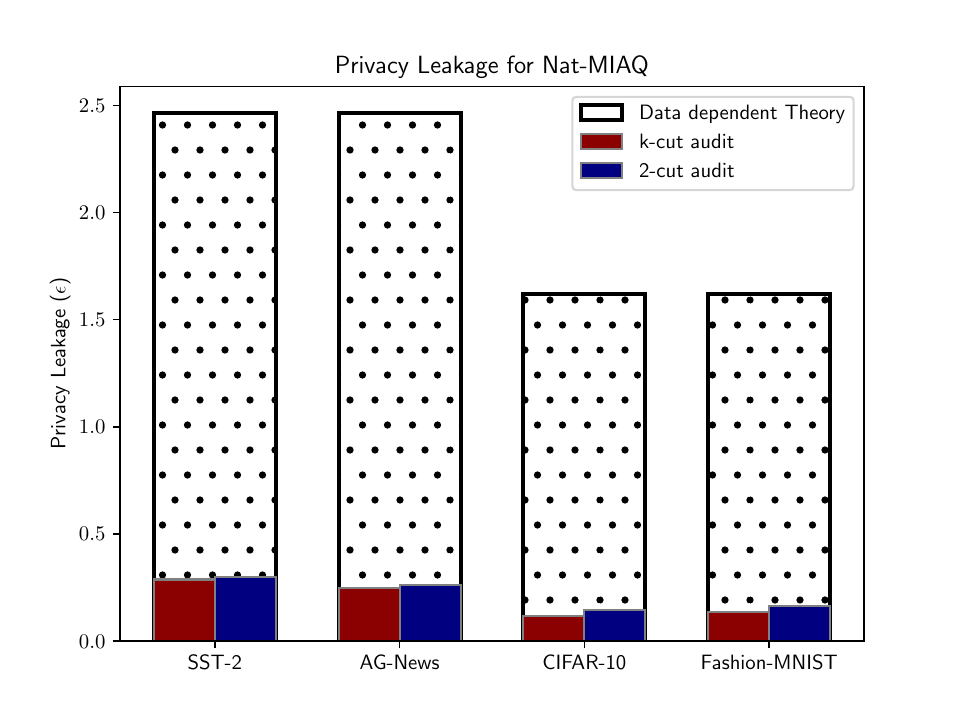}
        \caption{}
    \end{subfigure}%
    ~ 
    \begin{subfigure}[t]{0.45\textwidth}
        \centering
        \includegraphics[width=\textwidth]{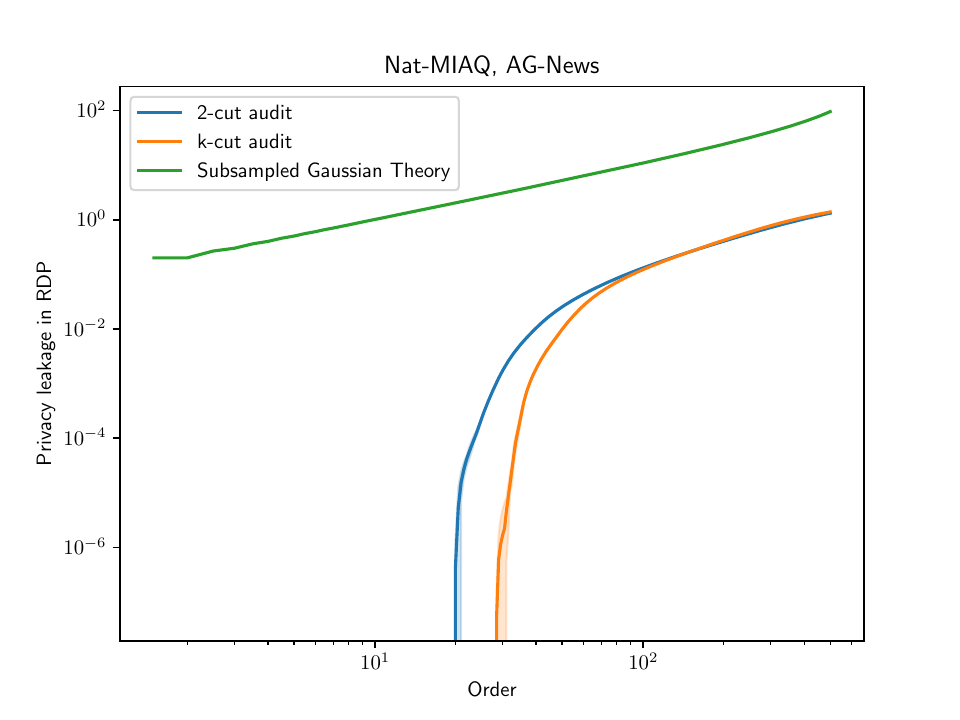}
        \caption{}
    \end{subfigure}
    \caption{Average case for $\natmiadvquer$ adversary}
\end{figure}

\begin{figure}[!htb]
    \centering
    \begin{subfigure}[t]{0.45\textwidth}
        \centering
        \includegraphics[width=\textwidth]{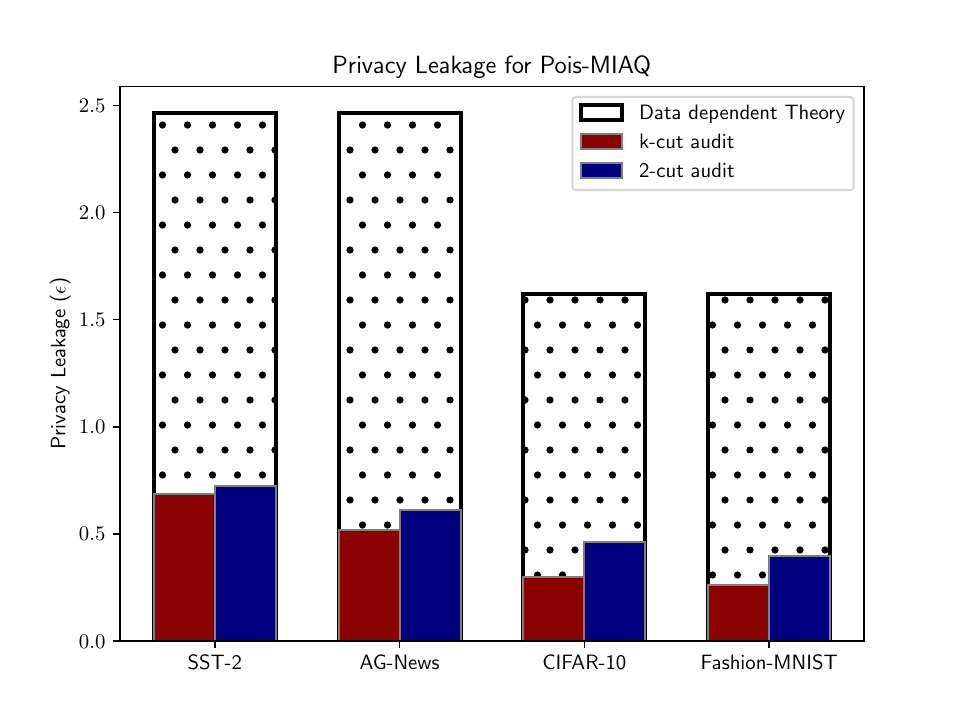}
        \caption{}
    \end{subfigure}%
    ~ 
    \begin{subfigure}[t]{0.45\textwidth}
        \centering
        \includegraphics[width=\textwidth]{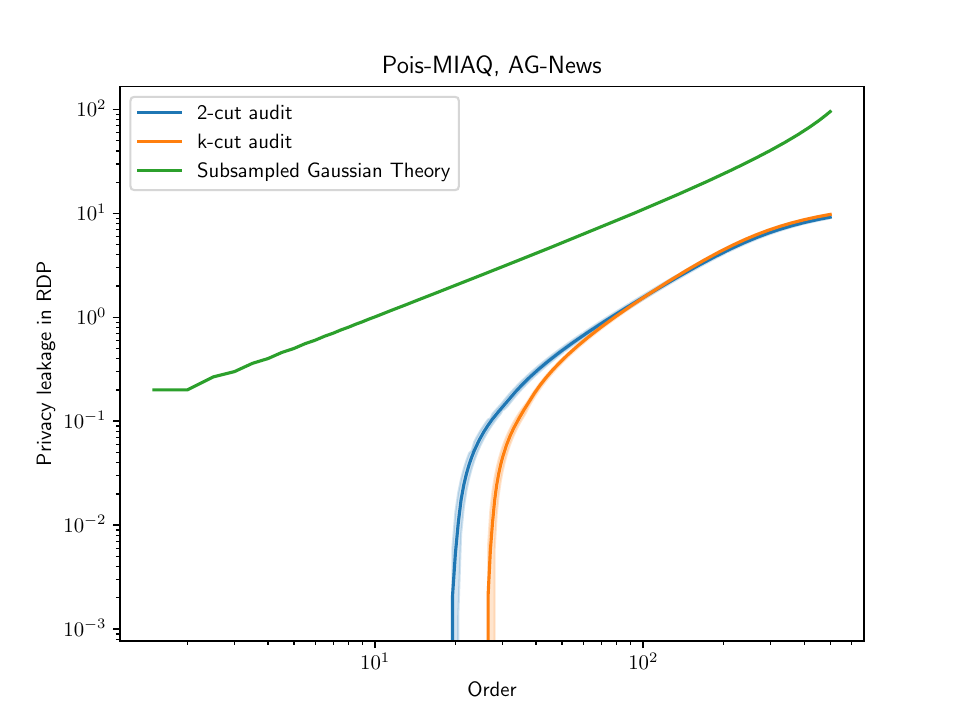}
        \caption{}
    \end{subfigure}
    \caption{Worst case for $\poismiadvquer$ adversary}
\end{figure}

\begin{figure}[!htb]
    \centering
    \begin{subfigure}[t]{0.45\textwidth}
        \centering
        \includegraphics[width=\textwidth]{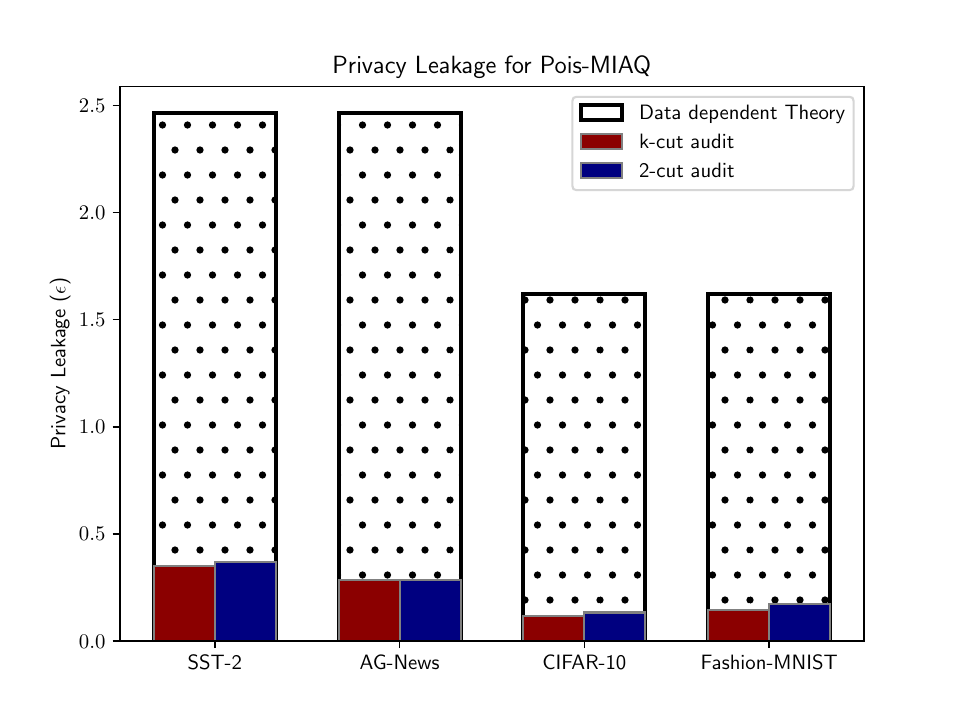}
        \caption{}
    \end{subfigure}%
    ~ 
    \begin{subfigure}[t]{0.45\textwidth}
        \centering
        \includegraphics[width=\textwidth]{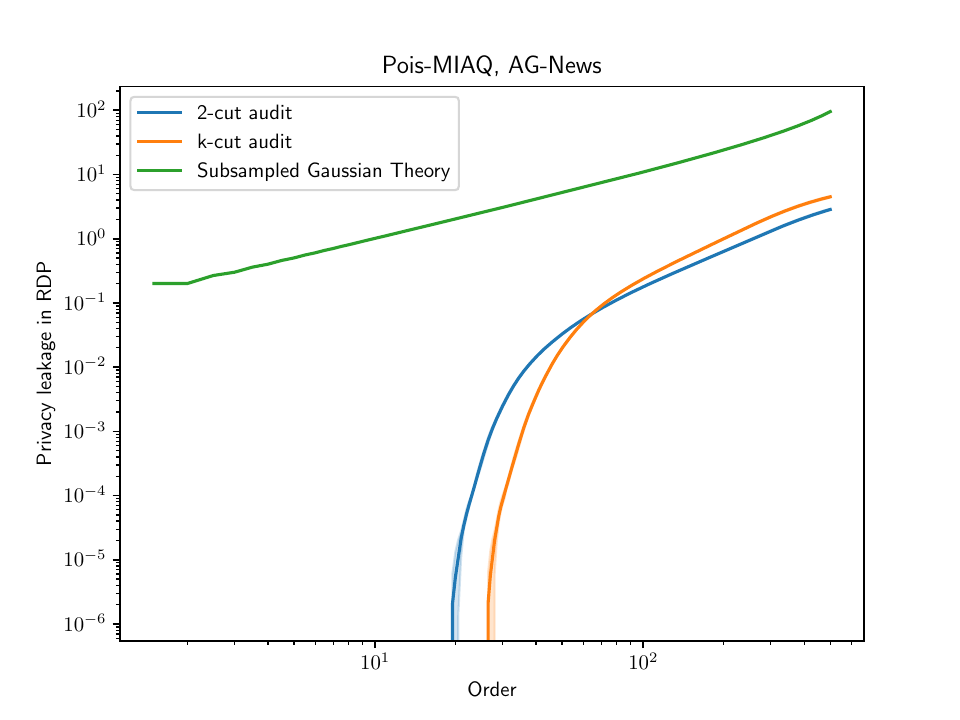}
        \caption{}
    \end{subfigure}
    \caption{Average case for $\poismiadvquer$ adversary}
\end{figure}

\begin{figure}[!htb]
    \centering
    \begin{subfigure}[t]{0.45\textwidth}
        \centering
        \includegraphics[width=\textwidth]{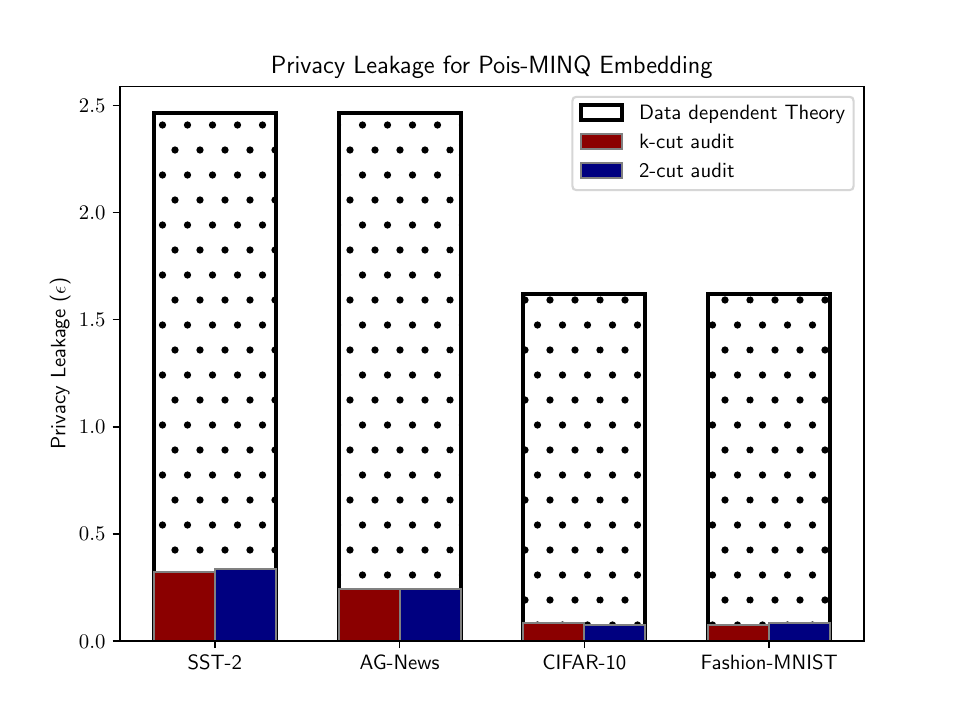}
        \caption{}
    \end{subfigure}%
    ~ 
    \begin{subfigure}[t]{0.45\textwidth}
        \centering
        \includegraphics[width=\textwidth]{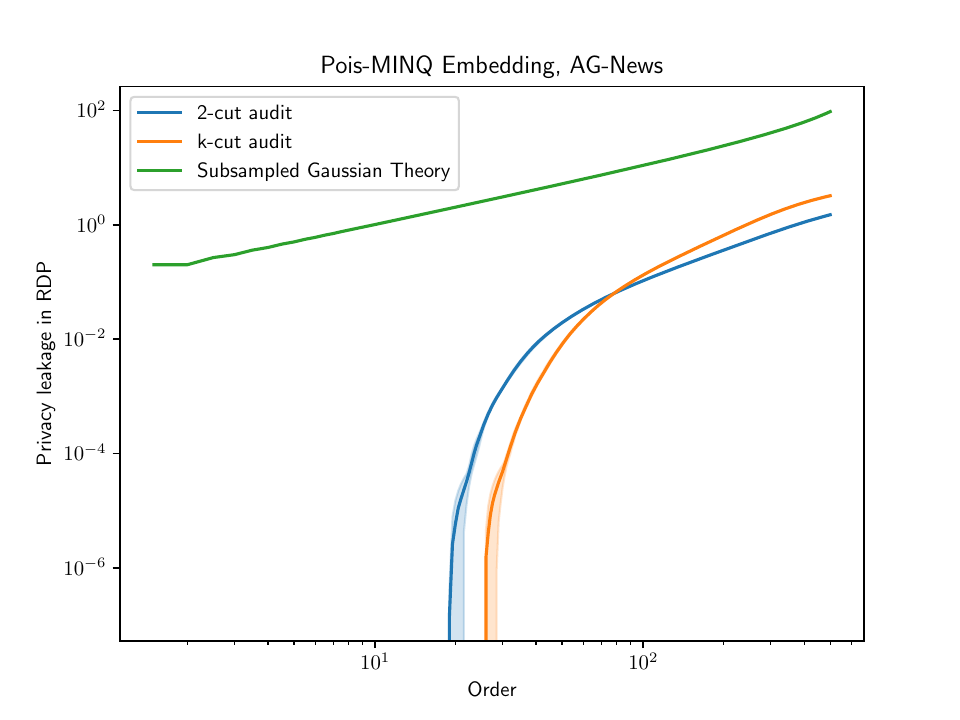}
        \caption{}
    \end{subfigure}
    \caption{$\poisminatquer$ Embedding adversary}
\end{figure}

\begin{figure}[!htb]
    \centering
    \begin{subfigure}[t]{0.45\textwidth}
        \centering
        \includegraphics[width=\textwidth]{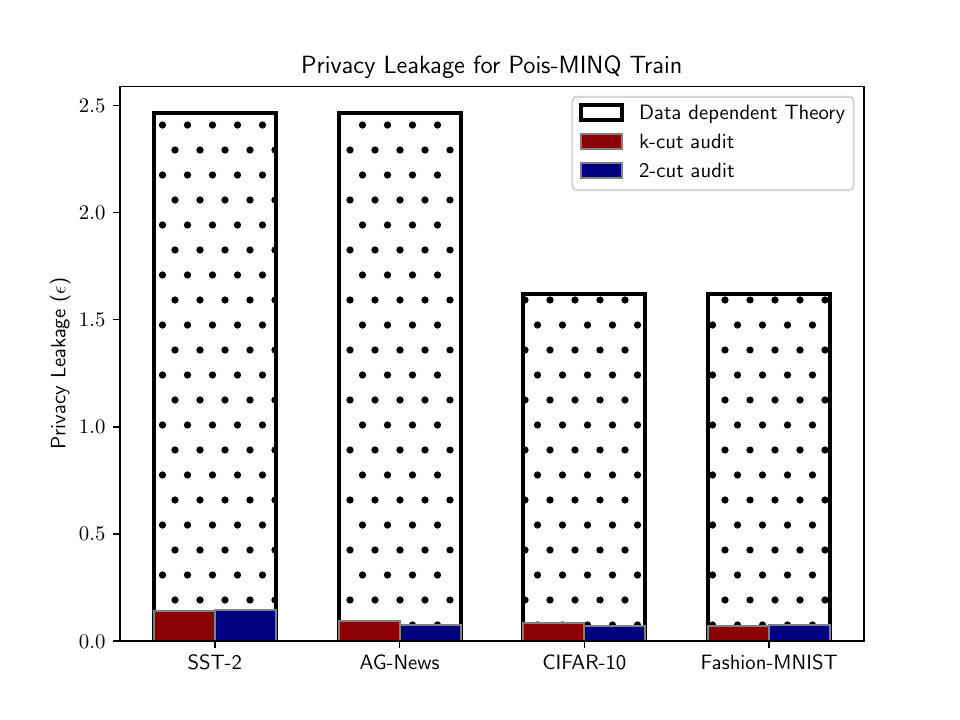}
        \caption{}
    \end{subfigure}%
    ~ 
    \begin{subfigure}[t]{0.45\textwidth}
        \centering
        \includegraphics[width=\textwidth]{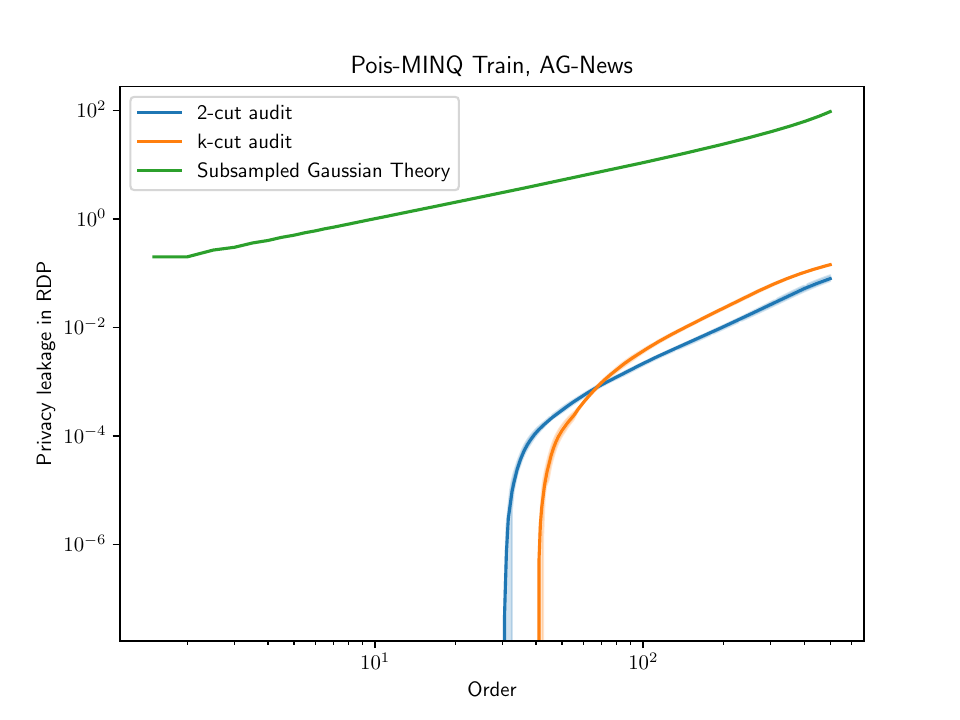}
        \caption{}
    \end{subfigure}
    \caption{$\poisminatquer$ Train adversary}
\end{figure}

\begin{figure}[!htb]
    \centering
    \begin{subfigure}[t]{0.45\textwidth}
        \centering
        \includegraphics[width=\textwidth]{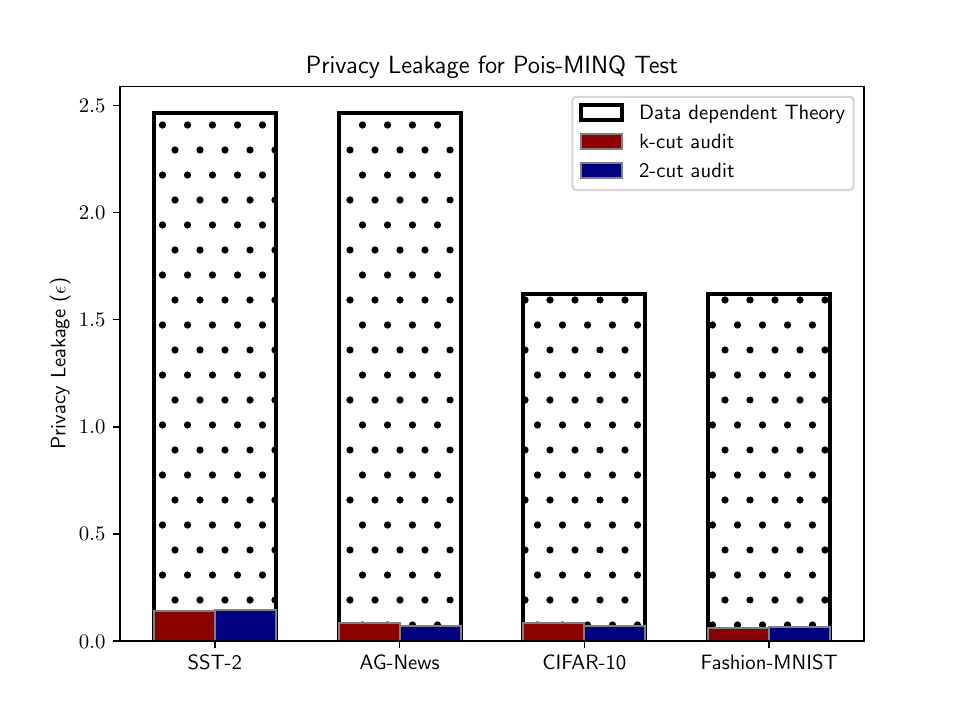}
        \caption{}
    \end{subfigure}%
    ~ 
    \begin{subfigure}[t]{0.45\textwidth}
        \centering
        \includegraphics[width=\textwidth]{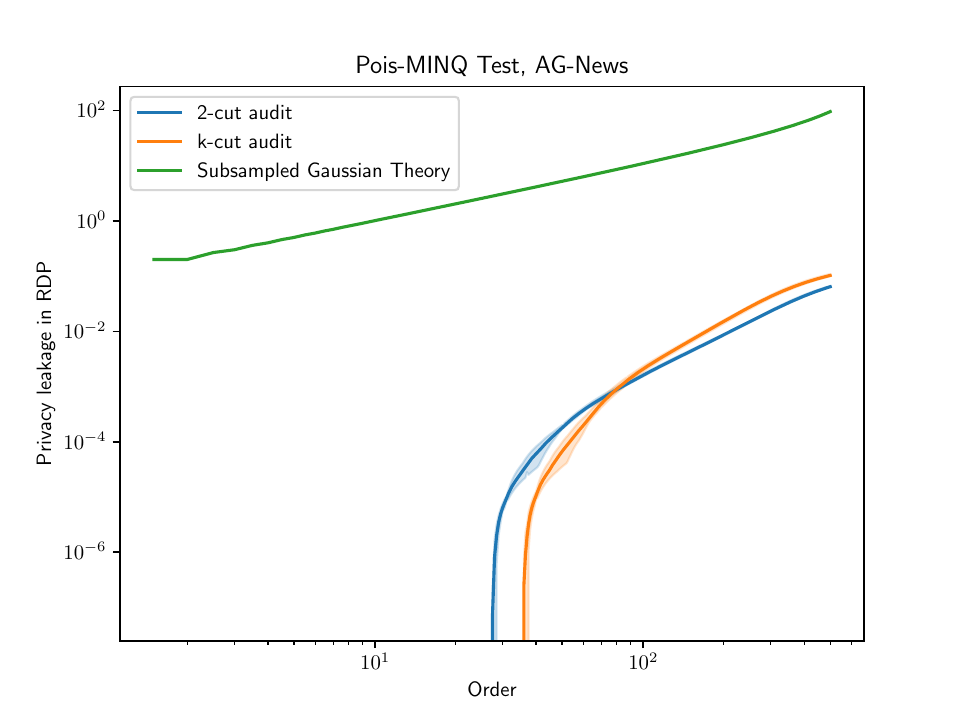}
        \caption{}
    \end{subfigure}
    \caption{$\poisminatquer$ Test adversary}
\end{figure}

\end{document}